\documentclass[twocolumn,showpacs,nofootinbib,preprintnumbers,amsmath,amssymb,showkeys]{revtex4}

\usepackage{epsfig}
\usepackage{color}% Include figure files
\usepackage{graphicx}% Include figure files
\usepackage{dcolumn}% Align table columns on decimal point
\usepackage{bm}% bold math
\usepackage{dsfont}
\newcommand{\beq}{\begin{equation}}
\newcommand{\eeq}{\end{equation}}
\newcommand{\bea}{\begin{eqnarray}}
\newcommand{\eea}{\end{eqnarray}}
\newcommand{\nn}{\nonumber \\}

\newcommand\eqn[1]{(\ref{#1})}      % parentheses around the LaTex "ref" macro
\newcommand\Eqn[1]{Eq.~(\ref{#1})}  % includes ``Eq.'' in front
\begin{document}

%\preprint{APS/123-QED}

\title{Yang-Mills correlators at finite temperature: A perturbative perspective}% Force line breaks with \\

\author{U. Reinosa}%
\affiliation{%
Centre de Physique Th\'eorique, Ecole Polytechnique, CNRS, 91128 Palaiseau Cedex, France
}%
\author{J. Serreau}%
\affiliation{%
 Astro-Particule et Cosmologie (APC), CNRS UMR 7164, Universit\'e Paris 7 - Denis Diderot\\ 10, rue Alice Domon et L\'eonie Duquet, 75205 Paris Cedex 13, France.
}%
\author{M. Tissier}
\affiliation{LPTMC, Laboratoire de Physique Th\'eorique de la Mati\`ere Condens\'ee, CNRS UMR 7600, Universit\'e Pierre et Marie Curie, \\ boite 121, 4 pl. Jussieu, 75252 Paris Cedex 05, France
}
\author{N. Wschebor}%
\affiliation{%
 Instituto de F\'{\i}sica, Facultad de Ingenier\'{\i}a, Universidad de la Rep\'ublica, J.H.y Reissig 565, 11000 Montevideo, Uruguay.
}%

\date{\today}% It is always \today, today,
             %  but any date may be explicitly specified

\begin{abstract}
We consider the two-point correlators of Yang-Mills theories at finite temperature in the Landau gauge. We employ a model for the corresponding Yang-Mills correlators
based on the inclusion of an effective mass term for gluons. The latter is expected to have its origin in the existence of Gribov copies. One-loop calculations at zero temperature have been shown to agree remarkably well with the corresponding lattice data. We extend on this and perform a one-loop calculation of the Matsubara gluon and ghost two-point correlators at finite temperature. We show that, as in the vacuum, an effective gluon mass accurately captures the dominant infrared physics for the magnetic gluon and ghost propagators. It also reproduces the gross qualitative features of the electric gluon propagator. In particular, we find a slight nonmonotonous behavior of the Debye mass as a function of temperature, however not as pronounced as in existing lattice results. A more quantitative description of the electric sector near the deconfinement phase transition certainly requires another physical ingredient sensitive to the order parameter of the transition. 
 \end{abstract}

% PACS, the Physics and Astronomy
                             % Classification Scheme.
%Use showkeys class option if keyword
                              %display desired
\maketitle

%\tableofcontents

%%%%%

\section{Introduction}
\label{sec:intro}

In the past two decades, an intense effort has been made in computing the
deep infrared behavior of the gluon and ghost correlators of
Yang-Mills theories in the vacuum, mostly in the Landau
 gauge, using either numerical simulations \cite{Boucaud:2011ug} or continuum approaches \cite{Alkofer00}. One important motivation goes back to the seminal works of Kugo and Ojima \cite{Kugo79} and Gribov and Zwanziger \cite{Gribov77,Zwanziger89,Zwanziger92}, who argued that the infrared behavior of such correlators might be a sensitive probe of confinement. A decade of detailed investigations has revealed that, at least in the Landau gauge, this relation is not as clear as originally expected. The other main motivation for such studies concerns the fact that, although these correlators are not directly observable, they are the basic building blocks for computing physical quantities using continuum approaches. It is thus crucial to control their behavior as precisely as possible. In this context, lattice simulations serve as a benchmark for various approximation schemes.

It is commonly accepted that the description of the infrared behavior of the vacuum ghost and gluon correlators requires nonperturbative methods since standard perturbation theory, based on the Faddeev-Popov construction, predicts an infrared Landau pole at low momentum. Powerful continuum methods have been developed following the seminal works of von Smekal, Alkofer and Hauck on Schwinger-Dyson equations \cite{vonSmekal97,Alkofer00} and of Ellwanger {\it et al.} on nonperturbative renormalization group techniques \cite{Ellwanger96}. In parallel to lattice simulations of ever increasing precision \cite{Cucchieri_08b,Cucchieri_08,Bornyakov2008,Cucchieri09,Bogolubsky09,Bornyakov09,Dudal10}, this has triggered intense activity in trying to accurately describe the deep infrared behavior of such correlators \cite{Alkofer00,Maas:2011se}.  For
pure SU(2) and SU(3) Yang-Mills theories in the Landau gauge a general agreement has now emerged for
space-time dimensions $d=2,3$ and $4$. In particular, in $d=4$, one obtains the so-called decoupling behavior where
both the gluon correlator and the ghost dressing function are finite at vanishing
momentum \cite{Boucaud06,Aguilar07,Aguilar08,Boucaud08,Fischer08,RodriguezQuintero10}. This is in contrast with the expectations from the Kugo-Ojima and the original Gribov-Zwanziger scenarios which favor the so-called scaling solution: the gluon
propagator vanishes at zero momentum while the ghost dressing function diverges.

These results suggest that the most relevant infrared physics can be
captured by an effective gluon mass. Following this line of thought,
two of us have investigated the vacuum gluon and ghost two-point
correlators in a simple massive extension of Landau gauge Yang-Mills
theories, which is a particular case of the Curci-Ferrari model
\cite{Tissier:2010ts,Tissier:2011ey}. Remarkably, in this model, one can push
perturbation theory down to the deep infrared regime because the gluon mass
suppresses potentially dangerous long wavelength fluctuations. Most
strikingly, the Landau pole that hampers the infrared calculations
within the Faddeev-Popov formalism is not an issue anymore. One-loop
results for the ghost and gluon two-point correlators in the vacuum
were shown to be in remarkable quantitative agreement with lattice
results in the Landau gauge. Similar conclusions extend to the
one-loop calculation of the three-point correlators
\cite{Pelaez:2013cpa}, assessing the fact that the dominant infrared
fluctuations in the vacuum and in the Landau gauge are indeed
described by a simple effective gluon mass.

These results have been put on more solid theoretical grounds with the proposal of a new one-parameter family of Landau gauges in Refs. \cite{Serreau:2012cg,Serreau:2013ds}. This is based on taking a particular average over Gribov copies, which avoids the usual Neuberger zero problem \cite{Neuberger:1986vv} of the standard Faddeev-Popov construction; see also \cite{vonSmekal:2008en}. The gauge-fixing procedure produces an effective bare mass for gluons which has to be sent to zero at the end of any calculation, together with the (continuum) limit of vanishing bare coupling. This can be done by keeping the renormalized mass finite, in which case the gauged-fixed action turns out to be perturbatively equivalent to the Curci-Ferrari model for what concerns the calculation of gluon and ghost correlators.

The investigation of the infrared properties of Yang-Mills correlators naturally extends to finite temperature, where direct numerical evaluations of the partition function using lattice techniques show a transition to a phase with deconfined gluonic excitations at a temperature $T_c$ \cite{Engels:1980ty,Boyd:1995zg}. The transition is of second order for the SU(2) group and of first order for SU(3).  As before, a detailed understanding of the thermal modifications of the gluon and ghost correlators is a key ingredient for continuum approaches \cite{Braun10,Fister:2013bh,Fischer12,Haas:2013hpa}. In that case, however, the status of gauge-fixed lattice calculations is not fully settled. Several
groups have performed lattice simulations of the finite temperature Matsubara 
correlators for the SU(2) and SU(3) Yang-Mills theories in $d=4$ \cite{Heller95,Heller97,Cucchieri00,Cucchieri01,Cucchieri07,Fischer10,Cucchieri11,Cucchieri12,Aouane:2011fv,Maas:2011ez,Silva:2013maa}.\footnote{For similar studies including dynamical quarks, see \cite{Furui06,Fischer10,Aouane12,Bornyakov11}.} Although at the quantitative level the lattice data still present
strong systematic uncertainties, in particular near the transition
temperature, some qualitative features emerge that seem rather robust (we focus on the correlators at zero Matsubara frequency):
\begin{itemize}
\item The ghost correlator is
  rather insensitive to thermal effects for temperatures up to 500~MeV
  \cite{Cucchieri07}.
\item The magnetic component of the gluon correlator depends smoothly on the temperature. The magnetic mass, defined as the square root of the inverse correlator at zero momentum, remains finite and increases monotonously with
  temperature. In the extreme infrared, the magnetic correlator is
  found to increase linearly with momentum, a behavior
  characteristic of the zero temperature gluon propagator in Landau
  gauge in $d=3$ \cite{Cucchieri07}. For low momenta, the magnetic correlator shows a strong sensitivity to the choice of Gribov copy \cite{Bornyakov:2010nc,Aouane:2011fv}.
\item The gluon electric propagator is more sensitive to thermal effects. It is also the one which shows the strongest sensitivity to lattice artifacts and is, to date, the less under control. The associated Debye mass has a nonmonotonous behavior
  as a function of temperature. It shows a minimum at a temperature roughly of the order of the transition temperature $T_c$. The precise position of the minimum and the actual ratio between the zero temperature mass and
  its smallest value around $T_c$ seem strongly sensitive to discretization effects \cite{Cucchieri11,Mendes:2014gva}. A comparison between the SU($2$) and the SU($3$) cases within comparable range of lattice parameters \cite{Fischer10} shows only slight quantitative differences, with a minimum at $T_c$ and a mass ratio of about 2. A more recent study focusing on the transition region in the SU($3$) case \cite{Aouane:2011fv} with, in particular, a careful extrapolation to the continuum limit finds a minimum at $T/T_c\approx0.86$ and a milder mass ratio of about 1.1.\footnote{It is to be emphasized though  that the results of \cite{Aouane:2011fv} are restricted to small physical volumes, which may bias the
determination of infrared properties such as the Debye mass.} The electric correlator is found to be essentially insensitive to the choice of Gribov copy \cite{Bornyakov:2010nc,Aouane:2011fv}.
\end{itemize}

Continuum approaches have also been implemented to compute Yang-Mills correlators at
finite temperature and some of the qualitative features described above (such as the insensitivity of the ghost correlator to thermal effects) have been noticed \cite{Fister11,Fukushima:2013xsa,Huber:2013yqa}; for a review, see \cite{Maas:2011se}. 

A natural question to ask is whether the relevant infrared physics at finite temperature can, as in the vacuum, be captured by an effective gluon mass. The smooth behavior of both the ghost and the magnetic gluon propagators support this hypothesis while the nonmonotonic behavior of the electric gluon propagator and the very existence of a phase transition suggest, in contrast, that there may be other types of infrared effects specific to the finite temperature. 

Here, we study Yang-Mills correlators at finite temperature in the class of Landau gauges proposed in \cite{Serreau:2012cg}. The equivalence with the Curci-Ferrari model still holds at finite temperature and, motivated by the results of \cite{Tissier:2010ts,Tissier:2011ey,Pelaez:2013cpa} in the vacuum, we perform a one-loop calculation of the Matsubara gluon and ghost propagators in this framework. Our main aim is to investigate to what extent the infrared effects at finite temperature can be described by an effective gluon mass. Choosing a mass and a coupling constant independent of the temperature, our one-loop results show a monotonous behavior of the magnetic gluon propagator and a nonmonotonous behavior of the electric one, in qualitative agreement with lattice data, although the drop in the Debye mass is much less pronounced. However, the corresponding ghost dressing function depends strongly on the temperature and even becomes singular for high temperature, in clear disagreement with lattice results.

A possible way to cure this problem is to allow for a temperature dependence of the gluon mass and of the coupling constant. Such a temperature dependence can be thought of as due to possible renormalization group effects as well as the fact that, according to \cite{Serreau:2012cg}, the gluon mass is an effective way of sampling Gribov copies and that the relevant value corresponding to a given lattice simulation may well depend on temperature. Indeed, the temporal extent of the lattice varies as the inverse temperature and thus so is the actual algorithm for finding a given Gribov copy. However, we are not able to determine the temperature dependence of the parameters analytically. Therefore, for each temperature, we fit the ghost and magnetic gluon propagators -- for which lattice data are better under control -- against various lattice data, from which we obtain the optimal mass and coupling constant. We get very good fits at all temperatures, which indicates that, as announced, these correlators are well described by a simple effective gluon mass. In particular, the ghost dressing function is essentially independent of the temperature and we reproduce the  transition from a $d=4$ to a $d=3$ behavior of the magnetic gluon propagator at low momenta. We find that the fitted mass is essentially independent of the temperature while the fitted coupling shows a slight decrease with increasing temperature. The latter is what prevents the singular behavior of the ghost dressing function mentioned above. 

We then use the obtained values to compute the electric gluon propagator and, in particular, the temperature dependence of the Debye mass which we can compare to the corresponding lattice data. We obtain a nonmonotonous behavior in rough qualitative agreement with lattice data. The agreement is not quantitative though in particular near the transition temperature. This shows that, again, an effective gluon mass captures the dominant infrared behavior away from $T_c$ but that another physical ingredient is needed to quantitatively reproduce the behavior of the Debye mass in the vicinity of $T_c$. This is in line with the expectation that the latter provides a sensitive probe of the phase transition \cite{Aouane:2011fv,Maas:2011ez}. A more refined treatment would imply taking into account the effect of a possible order parameter of the transition, such as the Polyakov loop. For instance, functional renormalization group studies \cite{Braun:2007bx,Marhauser:2008fz,Fister:2013bh} suggest that one should indeed take into account a possible nontrivial $A_0$ background, see also \cite{Xu:2011ud}. We plan to implement this possibility in the present perturbative framework in a future work.

The paper is organized as follows. In Sec.~\ref{sec_framework} we present the theoretical framework of the present work and describe the explicit calculation of the finite temperature ghost and gluon self-energies at one loop in the Curci-Ferrari model. We discuss the results of this calculation for temperature-independent mass and coupling constant in Sec.~\ref{sec_oneloop}. In Sec. \ref{sec_fits}, we compare our one-loop results to lattice data by fitting the mass and coupling at each temperature. Sec.~\ref{sec_concl} resumes our conclusions. Various technical details of the one-loop calculation are presented in Appendixes \ref{appsec:thermal}-\ref{appsec:sketch}.

%%%%%
\section{Theoretical framework}
\label{sec_framework}

To motivate the present theoretical framework let us briefly summarize the results of \cite{Serreau:2012cg}; see also \cite{Serreau:2013ds}.
The Landau gauge condition extremizes the functional $F[A]=\frac 12 \int_x A_\mu^a A_\mu^a$ along the gauge orbit $A\to A^U$, with $U$ an arbitrary gauge transformation. The existence of multiple extrema gives rise to the Gribov ambiguity issue which one needs to resolve in order to completely specify the gauge. The gauge-fixing procedure proposed in \cite{Serreau:2012cg} amounts to sampling Gribov copies along each gauge orbit with a nonuniform weight governed by the functional $F[A^U]$. This introduces an effective mass term for the gluon which simply reflects the degeneracy lift of Gribov copies. Including an appropriate sign factor in the sampling procedure, according to the sign of the Faddeev-Popov determinant of each copy, this gauge-fixing can be formulated as a local action by means of standard ghost
and antighost fields as well as a Nakanishi-Lautrup field to enforce the Landau gauge condition $\partial_\mu A_\mu^U=0$. 

The sampling measure must be properly normalized such that gauge invariant observables are blind to the gauge-fixing procedure. This introduces a nontrivial normalizing denominator which can be treated by means of the replica method of disordered systems in statistical physics \cite{disorder}. This amounts to introducing replicated sets of ghost, antighost and Nakanishi-Lautrup fields as well as replicated SU($N$) matrix fields. Remarkably, perturbative diagrams involving closed loops of these additional fields exactly cancel against each other and the latter thus eventually decouple in the calculation of correlators of the usual gluon and ghost fields, their only role being to provide an effective tree-level mass for the gluon. This is actually a consequence of an extended (super)symmetry involving the replicated ghost, antighost and scalar fields. The replica method imposes one to eventually send the mentioned bare mass to zero. This limit must be taken together with the continuum limit, corresponding to zero bare coupling. When both 
limits are taken in an appropriate way, i.e., such that the renormalized mass and couplings are finite, the proposed gauge fixing turns out to be perturbatively equivalent to the Landau version of the Curci-Ferrari model for what concerns correlation functions of gluons and ghosts. {  We are however not able so far to determine the value of this mass {\it ab initio} and we thus have two free parameters to adjust, one of which is related to the gauge-fixing procedure.}

At finite temperature, despite their anticommuting or commuting character, both types of fields have the same kind of (periodic) boundary conditions in the imaginary time direction and the corresponding supersymmetry is left unbroken. It follows that, for what concerns the calculation of gluon and ghost correlators, the gauge-fixed action effectively reduces to that the Curci-Ferrari model in the Landau limit:
\beq
\label{eq_CF}
 S=\!\int_x\left\{{1\over4}F_{\mu\nu}^aF_{\mu\nu}^a\!+\!{m^2\over2}A_\mu^aA_\mu^a\!+\!\partial_\mu\bar c^a(D_\mu c)^a\!+\!ih^a\partial_\mu A_\mu^a\right\}  ,
\eeq
with $\int_x=\int_0^\beta d\tau\int d^{d-1} x$ and
\begin{align}
 F_{\mu\nu}^a&=\partial_\mu A_\nu^a-\partial_\nu A_\mu^a+g\mu^\epsilon f^{abc}A_\mu^b A_\nu^c,\\
 (D_\mu c)^a&=\partial_\mu c^a+g\mu^\epsilon f^{abc}A_\mu^bc^c.
\end{align}
Here $d=4-2\epsilon$ is the space-time dimension, $\beta$ is the
inverse temperature and $\mu$ is an arbitrary mass scale introduced
such that the coupling constant $g$ is dimensionless. 

Let us discuss how the effective mass term in \eqn{eq_CF} may influence
 perturbative calculations at finite temperature. In the zero mass case, i.e. for the standard Faddeev-Popov action, there appear several momentum scales which are well separated in the high temperature limit, where the running coupling is sufficiently small \cite{Blaizot:2001nr}. The scale $T$ characterizes the typical excitations which contribute to bulk thermodynamic quantities such as the pressure or the entropy. Collective phenomena responsible, e.g., for the screening of long wavelength chromoelectric fields are characterized by the Debye mass $M_{D}\sim gT$. This scale sets the limit of validity of strict perturbation theory (i.e. without resummations): for momenta of this order, tree-level and loop contributions are of comparable size and one needs to employ resummation schemes, such as the hard thermal loop effective theory \cite{Braaten:1989mz,Frenkel:1989br,Blaizot:2001nr}. Finally, the magnetic mass $M_{\rm mag}\sim g^2T$ sets the scale for intrinsically nonperturbative dynamics  due to strong infrared effects \cite{Linde:1980ts,Blaizot:2001nr}. In fact the dynamical generation of a magnetic mass itself cannot be captured by standard perturbation theory.

The presence of the additional mass scale $m$ is susceptible to improving the convergence of perturbation theory when $m\gtrsim gT$ and would regulate strong infrared fluctuations for $m\gtrsim g^2T$. It clearly plays no role below this limit and we thus expect to face the issues of standard perturbation theory described above in the very high temperature regime. 
In the present paper, we are interested in a much lower temperature range, up to about twice the critical temperature. In that case, the running coupling is not small and the various thermal scales mentioned above are not clearly separated. Moreover the Curci-Ferrari mass modifies the usual perturbative expressions of the electric and magnetic masses in a nontrivial way. In particular, it gives rise to a nonzero magnetic mass already at one-loop order. In the range of temperatures considered here, we have $M_D\sim M_{\rm mag}$ and both masses are typically of the same order or slightly larger than $m$ as the temperature increases.\footnote{These qualitative arguments ignore a possible running of the mass with temperature.} For temperatures such that $M_D\sim M_{\rm mag}\gtrsim m$, we face again some of the issues of the standard perturbation theory and we expect the low momentum part of the propagators $p\lesssim M_D\sim M_{\rm mag}$ to receive sizable higher loop contributions.

%%%
\subsection{Two-point correlators}

We write the ghost self-energy in frequency-momentum space as $\Sigma^{ab}(K)=g^2N\delta^{ab}\Sigma(K)$, with $K=(\omega,\vec k)$. The ghost two-point function reads ${\cal G}^{ab}(K)=\delta^{ab}{\cal G}(K)$, with
\beq
 {\cal G}(K)=\frac{1}{K^2+g^2N\,\Sigma(K)}.
\eeq
Similarly, we write the gluon polarization tensor $\Pi_{\mu\nu}^{ab}(K)=g^2N\delta^{ab}\Pi_{\mu\nu}(K)$. In Landau gauge, the gluon two-point function  ${\cal G}_{\mu\nu}^{ab}(K)=\delta^{ab}{\cal G}_{\mu\nu}(K)$ is transverse with respect to $K$: $K_\mu{\cal G}_{\mu\nu}(K)=0$. At finite temperature, it thus admits the general Lorentz decomposition
\beq
 {\cal G}_{\mu\nu}(K)=P_{\mu\nu}^T(K){\cal G}_T(K)+P_{\mu\nu}^L(K){\cal G}_L(K),
\eeq
with the projectors
\beq
\label{eq:projTllll}
 P_{\mu\nu}^T(K)=(1-\delta_{\mu0})(1-\delta_{\nu0})\left(\delta_{\mu\nu}-{K_\mu K_\nu\over k^2}\right)
\eeq
and
\beq
\label{eq:projLperp}
 P_{\mu\nu}^T(K)+P_{\mu\nu}^L(K)=P_{\mu\nu}^\perp(K)=\delta_{\mu\nu}-{K_\mu K_\nu\over K^2}.
\eeq
The transverse ($T$) and longitudinal ($L$) components read
\beq
 {\cal G}_{T,L}(K)=\frac{1}{K^2+m^2+g^2N\,\Pi_{T,L}(K)},
\eeq
where 
\bea
\label{eq:projT}
 \Pi_{T}(K)&=&\frac{P^{T}_{\mu\nu}(K)\Pi_{\mu\nu}(K)}{d-2}\\
\label{eq:projL}
 \Pi_{L}(K)&=&P^{L}_{\mu\nu}(K)\Pi_{\mu\nu}(K).
\eea

The free propagators read ${\cal G}^{\rm free}(K)=G_0(K)$ and ${\cal G}^{\rm free}_{T}(K)={\cal G}^{\rm free}_{L}(K)=G_{m}(K)$, with
\beq
\label{eq:freeprop}
 G_\alpha(K)=\frac{1}{K^2+\alpha^2}.
\eeq
It sometimes proves useful to decompose a quantity $X$ as its $T=0$, vacuum part $X_{\rm vac}$
plus a thermal contribution that we denote by $X_{\rm th}$. 

%%%
\subsection{Ghost self-energy}
\label{sec:ghostself}

\begin{figure}[t!]  
\epsfig{file=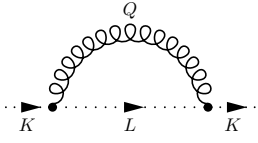,width=4.2cm}
 \caption{\label{fig:ghost} The ghost self-energy at one loop.}
\end{figure}

In this section we compute the one-loop contribution to the ghost self-energy depicted in Fig. \ref{fig:ghost}. We note $K\equiv(\omega,\vec k)$ the external momentum, $Q\equiv (\omega_n,\vec q)$ the internal loop momentum, with the bosonic Matsubara frequency $\omega_n\equiv 2\pi n T$, $n\in\mathbb{Z}$ and $L=K-Q$. Finally we use the notation
\beq
\int_Q f(Q)\equiv \mu^{2\epsilon}\,T\sum_{n\in\mathds{Z}}\int \frac{d^{d-1}q}{(2\pi)^{d-1}}f(\omega_n,\vec q).
\eeq
Below we show that the ghost self-energy can be entirely expressed in terms of the following elementary sum-integrals
\bea
\label{eq:int0}
J^\alpha & = & \int_Q G_\alpha(Q),\\
\label{eq:int1}
I^{\alpha\beta}(K) & = &\int_Q G_\alpha(Q) G_\beta(L),
\eea
where the indices $\alpha$ and $\beta$ refer to either the massless ($\alpha=0$) or massive ($\alpha=m$) propagators, \Eqn{eq:freeprop}. Notice the symmetry $I^{\alpha\beta}(K) =I^{\beta\alpha}(K)$.

The diagram of Fig. \ref{fig:ghost} reads
\beq
\Sigma(K)=- \int_Q \,P_{\mu\nu}^\perp(Q) L_\mu K_\nu\,G_{m}(Q)G_0(L),
\eeq
where the projector $P^\perp$ is defined in \eqn{eq:projLperp}. Writing
\beq
 P_{\mu\nu}^\perp(Q) L_\mu K_\nu=P_{\mu\nu}^\perp(Q) K_\mu K_\nu=K^2-\frac{(K\cdot Q)^2}{Q^2},
\eeq
one gets
\beq
\label{eq:sss}
\Sigma(K)= \int_Q \frac{(K\cdot Q)^2}{Q^2}\,G_{m}(Q)G_0(L)-K^2 I^{m0}(K),
\eeq
where each sum-integral on the right is well defined in dimensional regularization.

We eliminate the $1/Q^2$ term in the first sum-integral in \eqn{eq:sss} by using the identity ($\alpha\neq\beta$)
\beq
\label{eq:id1}
G_\alpha(Q)G_\beta(Q)=\frac{1}{\beta^2-\alpha^2}\big[G_\alpha(Q)-G_\beta(Q)\big].
\eeq
Furthermore, writing ($\alpha=0$ or $m$)
\beq
2K\cdot Q=K^2-\alpha^2+Q^2+\alpha^2-L^2,
\eeq
we obtain
\begin{align}
2 (K\cdot Q)\,G_\alpha(Q)G_0(L)&=\left(K^2-\alpha^2\right)G_\alpha(Q)G_0(L)\nn
&+G_0(L)-G_\alpha(Q),
\end{align}
and, applying the same trick twice,
\begin{align}
\label{eq:id2}
&4(K\cdot Q)^2G_\alpha(Q)G_0(L)={\left(K^2-\alpha^2\right)^2}G_\alpha(Q)G_0(L)\nn
&\qquad\quad+({K^2-\alpha^2+2K\cdot Q})\big[G_0(L)-G_\alpha(Q)\big].
\end{align}
Using the identities \eqn{eq:id1} and \eqn{eq:id2}, \Eqn{eq:sss} rewrites as 
\begin{align}
 \label{eq:ghostself}
 \Sigma(K)&=\frac{K^2-m^2}{4m^2}\left[J^m-J^0\right]+\frac{K^4}{4m^2}I^{00}(K)\nn
 &-\frac{(K^2+m^2)^2}{4m^2} I^{m0}(K)\,,
\end{align}
where we have used that $\int_Q (K\cdot Q)\big[G_0(Q)-G_m(Q)\big]=0$.
Thus, as announced, the ghost self-energy can be written in terms of simple well-known one-loop sum-integrals \eqn{eq:int0}-\eqn{eq:int1}, the calculation of which is recalled in Appendixes~\ref{appsec:thermal} and \ref{appsec:vac}.

\begin{figure}[t!]  
\epsfig{file=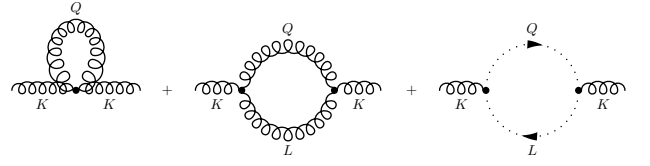,width=8.2cm}
 \caption{\label{fig:gluon} The gluon self-energy at one loop.}
\end{figure}

%%%
\subsection{Gluon self-energy}

The one-loop gluon self-energy involves the three diagrams of Fig. \ref{fig:gluon}, where, again, the internal gluon lines are massive. Using similar manipulations as in the previous section, it can be entirely expressed in terms of the elementary sum-integrals \eqn{eq:int0}-\eqn{eq:int1}, together with
\beq
\label{eq:int2}
I^{\alpha\beta}_{\mu\nu}(K) =  \int_Q Q_\mu Q_\nu\,G_\alpha(Q)G_\beta(L)=I^{\alpha\beta}_{\nu\mu}(K).
\eeq
Since only the transverse and longitudinal projections of the gluon polarization tensor, Eqs.~\eqn{eq:projT} and \eqn{eq:projL}, are needed to compute the gluon propagator in the Landau gauge, only the corresponding projected integrals are needed here:
\bea
I^{\alpha\beta}_{T}(K) & = & \frac{P^{T}_{\mu\nu}(K)I^{\alpha\beta}_{\mu\nu}(K)}{d-2},\label{appeq:ITdef}\\
I^{\alpha\beta}_{L}(K) & = & P^{L}_{\mu\nu}(K)I^{\alpha\beta}_{\mu\nu}(K).\label{appeq:ILdef}
\eea
Both have the symmetry property
\beq
\label{eq:symmetry}
 I_{T,L}^{\alpha\beta}(K)=I_{T,L}^{\beta\alpha}(K),
\eeq
which follows from (\ref{eq:int2}) and
\beq
 P^{T,L}_{\mu\nu}(K)Q_\mu Q_\nu=P^{T,L}_{\mu\nu}(K)L_\mu L_\nu.
\eeq

It is straightforward, albeit too long to be detailed here, to show that
\begin{widetext}
\bea
 \Pi_{T,L}(K) &=& \left(1-\frac{K^4}{2m^4}\right) \!I_{{ T,L}}^{00}(K)+ \left(1+\frac{K^2}{m^2}\right)^{\!\!2}\!I_{{ T,L}}^{m0}(K)-2\left[d-2+\left(1+\frac{K^2}{2m^2}\right)^{\!\!2}\right] \!I^{mm}_{ T,L}(K)\nonumber\\
 \label{eq:gluonself}
&+&(d-2)J^m-\left(K^2\!+\!m^2\right)\,I^{m0}(0)+\frac{\left(K^2\!+\!m^2\right)^2}{m^2}\,I^{m0}(K)-K^2\!\left(4\!+\!\frac{K^2}{m^2}\right)\! I^{mm}(K).\nn
\eea
\end{widetext}
A sketch of the steps leading to these expressions is given in Appendix \ref{appsec:sketch}.

%%%
\subsection{Renormalization}

All the sum-integrals appearing in Eqs. \eqn{eq:ghostself} and \eqn{eq:gluonself} can be decomposed as the sum of vacuum and thermal contributions. The thermal contributions are UV finite and can be expressed, after setting $d=4$, in terms of one-dimensional integrals, as detailed in Appendix \ref{appsec:thermal}.  The vacuum contributions can be computed analytically, as shown in Appendix~\ref{appsec:vac}, and contain UV divergences that need to be renormalized away. 

To this purpose we redefine $g^2N\Sigma(K)$ and $g^2N\Pi_{\mu\nu}(K)$ by adding the contributions $\delta Z_c K^2$ and $\delta m^2\delta_{\mu\nu}+\delta Z_A K^2P^\perp_{\mu\nu}(K)$ respectively and we adjust the counterterms $\delta Z_c$, $\delta m^2$ and $\delta Z_A$ such that
\beq
\label{eq:renormcond}
\Sigma_{\rm vac}(K^2=\mu^2)=\Pi_{\rm vac}(K^2=0)=\Pi_{\rm vac}(K^2=\mu^2)=0. 
\eeq
We have checked that the expressions for $\Sigma_{\rm vac}(K)$ and $\Pi_{\rm vac}(K)$ obtained from (\ref{eq:ghostself}) and (\ref{eq:gluonself}) coincide with those obtained in \cite{Tissier:2010ts,Tissier:2011ey}.
%\begin{align}
%\delta Z_A&=\frac{13}{6}\frac{g^2N}{16\pi^2\varepsilon} + ???????\\
%\delta Z_c&=\frac{3}{4}\frac{g^2N}{16\pi^2\varepsilon}+ ???????\\
%\frac{\delta {m^2}}{m^2}&=-\frac{3}{4}\frac{g^2N}{16\pi^2\varepsilon}+ ???????
%\end{align}

%%%%%
\section{One-loop results}
\label{sec_oneloop}
We now discuss our one-loop results for various quantities of interest for fixed values of the mass and coupling parameters. 
We use the values $m=0.68$~GeV and $g=7.5$ defined at the renormalization scale $\mu=1$~GeV, which were found to
give the best fits for the gluon and ghost propagators at $T=0$ for
the SU(2) group (see Sec. III B of \cite{Tissier:2011ey}). 

%%%
\subsection{Debye mass}
The behavior of the Debye mass is easily extracted from the behavior
of the gluon self-energy for small momentum:
\beq
M_D^2(T)=m^2+g^2N\,\Pi_L(0,k\to0).
\eeq

In the renormalization scheme considered here, the
zero-temperature contribution to the Debye mass is exactly
$m^2$. Moreover, the thermal contribution has a simple analytic expression which can be obtained from (\ref{eq:gluonself}) and the explicit expressions for $I^{\alpha\beta}_{L,{\rm th}}(0,k\to 0)$ and $I^{\alpha\beta}_{T,{\rm th}}(0,k\to 0)$ given in Appendix~\ref{appsec:vac}. Altogether,
\begin{align}
 \label{eq:AAA}
 M_D^2(T)&=m^2-g^2N\bigg[\frac{T^2}{24}\left(1+{4\pi^2\over5}{T^2\over m^2}\right)\nn
 &-{m^2\over2\pi^2}\int_0^\infty\!\!dq\,{n_{\varepsilon_q}\over\varepsilon_q}\left(3+{6q^2\over m^2}+{q^4\over m^4}\right)\bigg].
\end{align}

\begin{figure}[ht]
  \centering
\includegraphics[width=.95\linewidth]{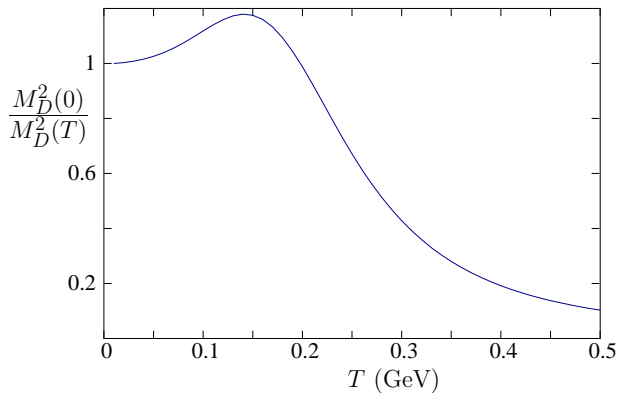}
  \caption{Inverse Debye mass normalized to its $T=0$ value, as a function of temperature. The curve shows a nonmonotonous behavior with a maximum at a temperature of the order of 0.14 GeV.}
  \label{fig_debye}
\end{figure}

As shown in Fig. \ref{fig_debye}, our expression reproduces the
nonmonotonous behavior of the Debye mass. The amplitude of the
variation is however not as large as that observed in lattice
simulations, where the ratio $M^2_D(0)/M^2_D(T)$ can be as large
as 4 near $T=T_c$ \cite{Maas:2011ez,Silva:2013maa}. Recall however that the lattice results
show a very large dependence both on the volume and on the lattice
spacing \cite{Cucchieri11,Cucchieri12}. The
position of the peak is found here to be at  $T_{\rm peak}\approx140$~MeV, while, in
lattice simulations, it is found to be close to $T_c\sim
300$~MeV.  It is interesting to discuss the dependence of the Debye mass with the parameters $m$ and $g$. One easily sees using \Eqn{eq:AAA} that the position of the peak is completely determined by the value of $m$ whereas its height relative to the zero temperature value of the peak is controlled by $g$. More precisely
\beq
\label{eq:peak}
\frac{T_{\rm peak}}{m}\simeq 0.2
\qquad
{\rm and} 
\qquad
\frac{M^2_D(0)}{M^2_D(T_{\rm peak})}\simeq\frac{1}{1-0.0014\,g^2N}\,.
\eeq

In the limit $m\ll T$, one recovers the standard one-loop result
\beq
M_D^2(T)\sim\frac{g^2N }{3}T^2\equiv M^2_{D,\infty}(T).\label{eq_asym_debye}
\eeq

\begin{figure}[ht]
  \centering
\includegraphics[width=.95\linewidth]{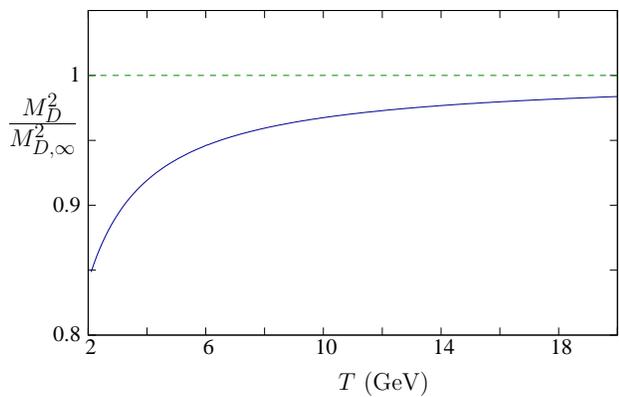}
\caption{Debye mass divided by its asymptotic form $M_{D,\infty}$ [see
  Eq. (\ref{eq_asym_debye})] as a function of temperature (plain
  curve). The deviation is significant up to rather large
  temperatures.}
\label{fig_debye_mass_over_T2}
\end{figure}
The convergence toward the asymptotic result is rather slow, as
depicted in Fig. \ref{fig_debye_mass_over_T2}. This means that the
effect of the mass $m$ remains sizeable even at temperatures much
larger than $T_c$.

%%%
\subsection{Magnetic mass}

A similar study can be performed in the magnetic sector. 
Introducing the magnetic mass as
\beq
M_{\text {mag}}^2(T)=m^2+g^2N\,\Pi_T(0,k\to0),
\eeq
we find the following analytic expression
\begin{align}
  \label{eq_mag_mass}
  M_{\text {mag}}^2(T)&=m^2+g^2 N\bigg[\frac{T^2}{24}\left(1+\frac{4\pi^2} {15}{T^2\over m^2}\right)\nn
   &-\frac {m^2}{6\pi^2}\int_0^\infty\!\!dq\,{n_{\varepsilon_q}\over\varepsilon_q}\left(3\frac{q^2}{m^2} + \frac{q^4}{ m^4}\right)\bigg].
\end{align}

\begin{figure}[ht]
  \centering
\includegraphics[width=.95\linewidth]{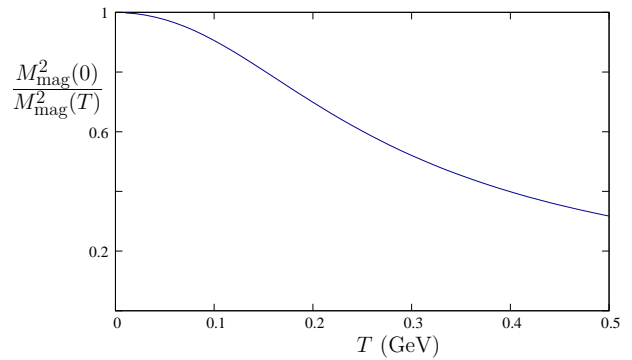}
\caption{Inverse magnetic mass normalized to its $T=0$ value, as a
  function of temperature. The variation of the magnetic mass is less
  dramatic than that of the Debye mass.}
  \label{fig_mag_mass}
\end{figure}

The magnetic mass is monotonous as a function of temperature such that $M_{\rm mag}(T)\ge m$; see
Fig.~\ref{fig_mag_mass}.
This is in agreement with lattice results.

The asymptotic behavior $m\ll T$ is qualitatively different from that
of the Debye mass, the
increase of the magnetic mass with the temperature being only linear:
\begin{equation}
  \label{eq_asym_mag_mas}
  M_{\text  {mag}}^2(T)=\frac {g^2N}{6\pi} m T+\mathcal O(m^2).
\end{equation}
This calculation has been done within perturbation theory at
one loop. However, the one-loop correction becomes of the order of the
tree-level contribution when $m\sim g^2 N T/(6\pi)$. For higher
temperatures, it becomes necessary to resum higher loop corrections in
the spirit of hard thermal loops. This would give a $T$-dependent
magnetic mass that would grow as:
\begin{equation}
  \label{eq:mmassasym}
  M_{\text  {mag}}(T)\stackrel{T\gg m}{\sim} \frac{g^2 N}{6\pi} T.
\end{equation}
As for the Debye mass, the approach to the asymptotic behavior is
rather slow. It is interesting to notice that for small temperatures,
the Debye mass is actually smaller than the magnetic mass, at odds
with the behavior expected at high temperatures but in agreement with
lattice results.

%%%
\subsection{Ghost dressing function}

The ghost dressing function is defined as usual as
\begin{equation}
  \label{eq_def_dressing}
  F(K)=K^2\mathcal G(K).
\end{equation}
It can be obtained in closed form at vanishing momentum in a similar way as above. Using the results of Appendix \ref{appsec:vac}, we obtain
\begin{align}
\label{eq:F0}
F^{-1}(0) &= F^{-1}_{\rm vac}(0)+g^2N\bigg[-\frac{1}{24}\frac{T^2}{m^2}\left(1+\frac{4\pi^2} {15}{T^2\over m^2}\right)\nn
&+\frac{1}{6\pi^2}\int_0^\infty dq\,\frac{n_{\varepsilon_q}}{\varepsilon_q}\left(3\frac{q^2}{m^2}+\frac{q^4}{m^4}\right)\bigg].
\end{align}
This analytic expression is however not
independent of the previous one. Indeed, there is a nonrenormalization
theorem that relates the magnetic propagator and the ghost dressing
function at vanishing momentum:
\begin{equation}
\label{eq_nonren}
  \left.\mathcal G_{T,B}^{-1}(K)F_B^{-1}(K)\right|_{\omega=0,k\to0} =m_B^2
\end{equation}
where the index $B$ denotes bare quantities. This theorem is a straightforward
generalization of the corresponding one for
$T=0$ \cite{Dudal02,Wschebor:2007vh,Tissier:2010ts,Tissier:2011ey} \footnote{The proof is based
  on the Slavnov-Taylor identity for the two-point function taken at
  small four-momentum; see, e.g., Appendix A of Ref.~\cite{Tissier:2011ey}. Since the frequencies are discrete at finite
  temperature, the frequency must be taken strictly to zero so that
  the electric sector is not constrained.} At one-loop order, this leads to a constraint for the thermal
part of (\ref{eq_nonren}) which reads
\begin{align}
  g^2N\,\Sigma_{\rm th}(0,k\to0)&\sim-\frac{k^2}{m^2}g^2N\,\Pi_{T,{\rm th}}(0,k\to0)\nn&\sim k^2\left(1-\frac{M^2_{\rm mag}(T)}{m^2}\right)\le0,
\end{align}
which yields (\ref{eq:F0}). It is a nontrivial check that our one-loop expression for $F(K)$ is
indeed consistent with this result.

\begin{figure}[ht]
  \centering
\includegraphics[width=.95\linewidth]{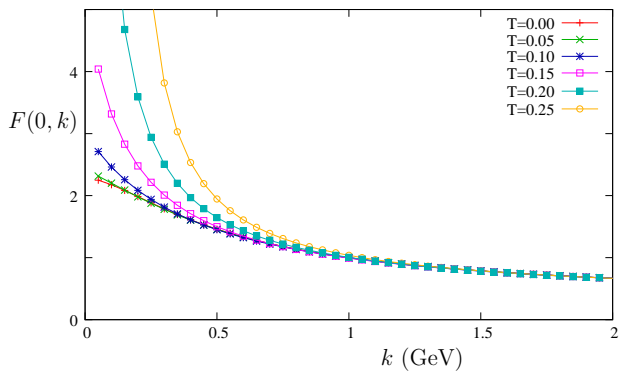}  
\caption{Ghost dressing function as a function of the momentum, for
  different temperatures, ranging from 0 to 0.25 GeV.  The infrared
  dressing function is strongly enhanced as the temperature is
  increased. It even shows a pole for temperatures larger than 0.25
  GeV.}
  \label{fig_ghost_dressing}
\end{figure}

In the high temperature limit, we find
\begin{equation}
  \label{eq:F0asym}
  F^{-1}(0,k\to0)\stackrel{T\gg m}{\sim}F^{-1}_{\rm vac}(0) +1-\frac{g^2N  }{6\pi
  }\,\frac T m
\end{equation}
Consequently, working at $m$ independent of the temperature, we are
doomed to hit a singularity for $F(0,k\to0)$ for sufficiently large
temperature. This can be seen on Fig.~\ref{fig_ghost_dressing} which shows the ghost dressing function as a function of momentum for various temperatures. Using the parameters extracted from the $T=0$
propagators, the pole appears at a temperature $T\sim
200$~MeV. We shall see below that this can be resolved by allowing for a dependence of the mass and coupling parameters on the temperature. 

%%%
\subsection{Gluon propagators}

\begin{figure}[ht]
  \centering
\includegraphics[width=.95\linewidth]{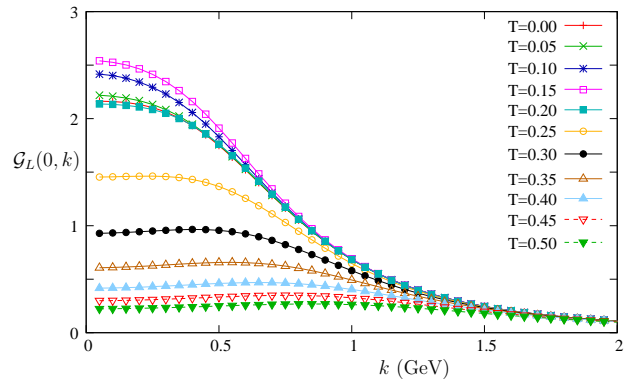}  
  \caption{Longitudinal propagator as a function of the momentum, for different temperatures, ranging from 0 to 0.5 GeV. The large momentum behavior is unaffected by the temperature. The curve for $T=0$ is almost degenerate with that for $T=0.2$~GeV.}
  \label{fig_long_propag}
\end{figure}

We now describe the behavior of the propagators. We present in
Fig.~\ref{fig_long_propag} the behavior of the longitudinal propagator
for different temperatures.
These propagators are relatively flat at low momenta. We also verify that, as expected, thermal effects only occur for modes roughly below the first Matsubara frequency $2\pi T$.

\begin{figure}[ht]
  \centering
\includegraphics[width=.95\linewidth]{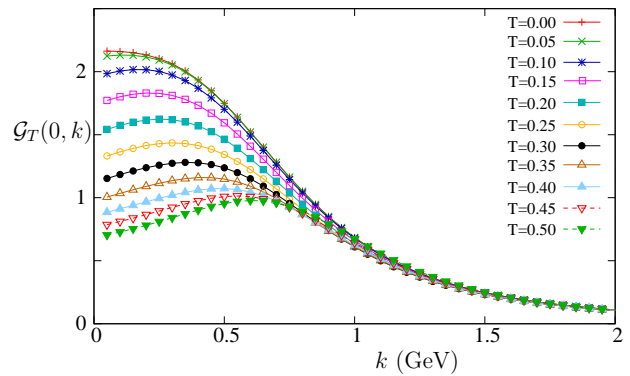}  
\caption{Transverse propagator as a function of the momentum, for
  different temperatures, ranging from 0 to 0.5 GeV. The linear
  infrared behavior is reminiscent of the $3d$-behavior of the gluon
  propagator at vanishing temperature.}
  \label{fig_trans_propag}
\end{figure}

The transverse propagators shown in Fig.~\ref{fig_trans_propag} have a
rather different behavior. For momenta larger than 800~MeV, it is
almost independent of the temperature. For smaller momenta, one
observes a linear growth as observed in the lattice results. This
linear growth is a characteristic feature of the three dimensional $T=0$ gluon
propagators in the Landau gauge. This is to be expected since, when
the temperature grows, one expects that the theory behaves like a $3d$
gauge theory coupled to a heavy scalar \cite{Appelquist:1981vg}.

\section{Comparison with lattice simulations}
\label{sec_fits}

The analysis of the previous section is done for fixed values of the parameters $m$ and $g$. 
However, in principle, the latter could depend on the temperature. For instance, in the present context, the bare mass is to be seen as an effective way to take into account the Gribov copies \cite{Serreau:2012cg,Serreau:2013ds}. It is thus expected to depend on $T$ since the Gribov copies of the $T=0$ and $T\neq 0$ are different. Moreover, it is well known that the resummation of
the hard thermal loops, which is mandatory at sufficiently large
temperatures, implies, among other things, the introduction of
$T$-dependent mass and coupling. 

In the present section, we allow for such a temperature dependence of the renormalized parameters $m$ and $g$. Since we do not know at present how to determine this dependence analytically, we choose to determine the values of these parameters at each temperature by fitting lattice data with our one-loop results. 

\begin{figure}[ht]
  \centering
  \includegraphics[width=.9\linewidth]{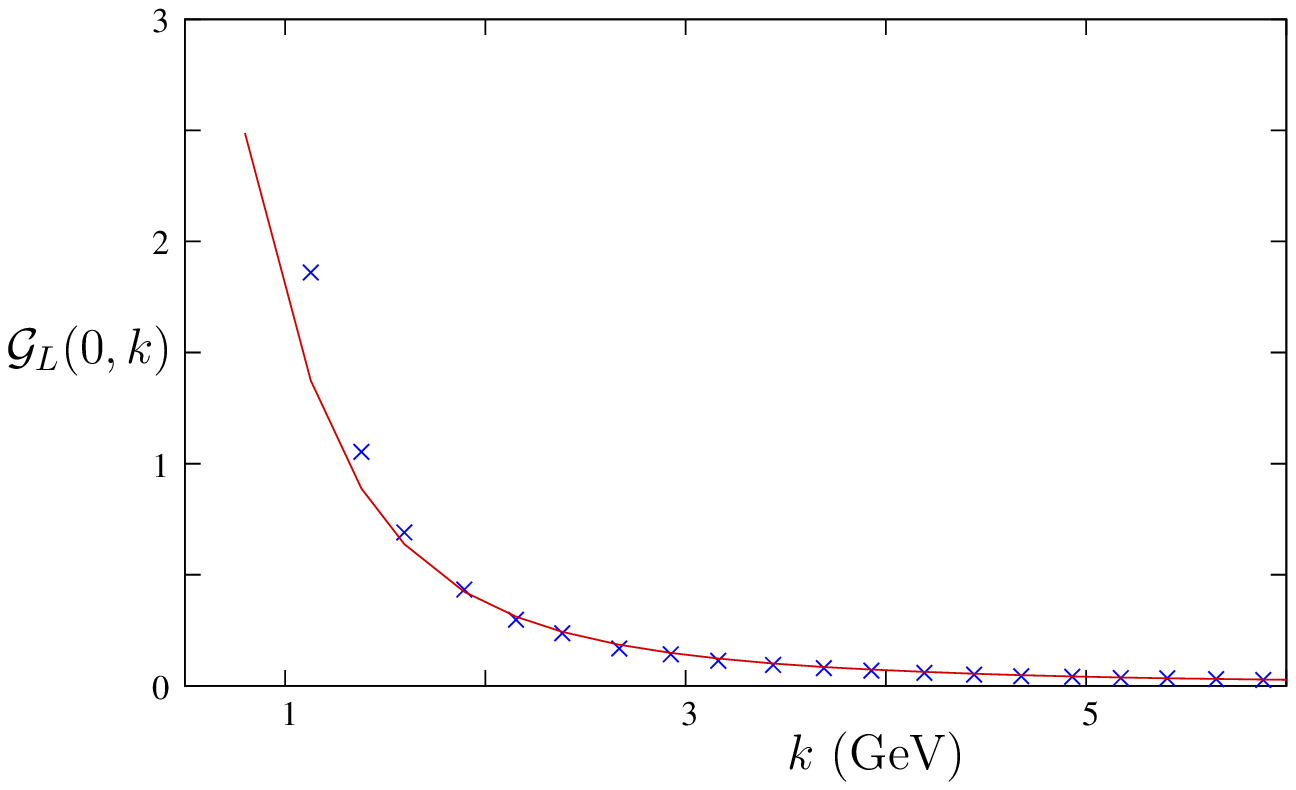}
  \includegraphics[width=.9\linewidth]{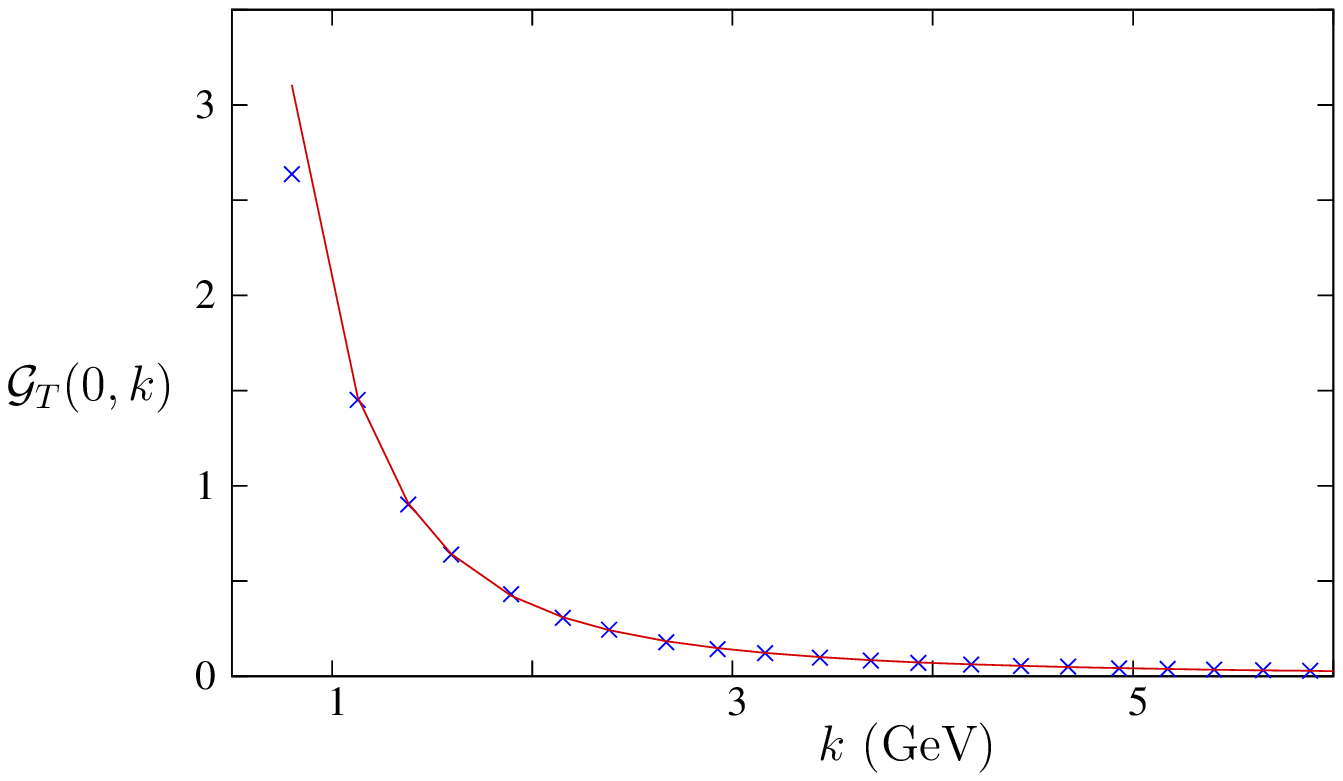}
  \includegraphics[width=.9\linewidth]{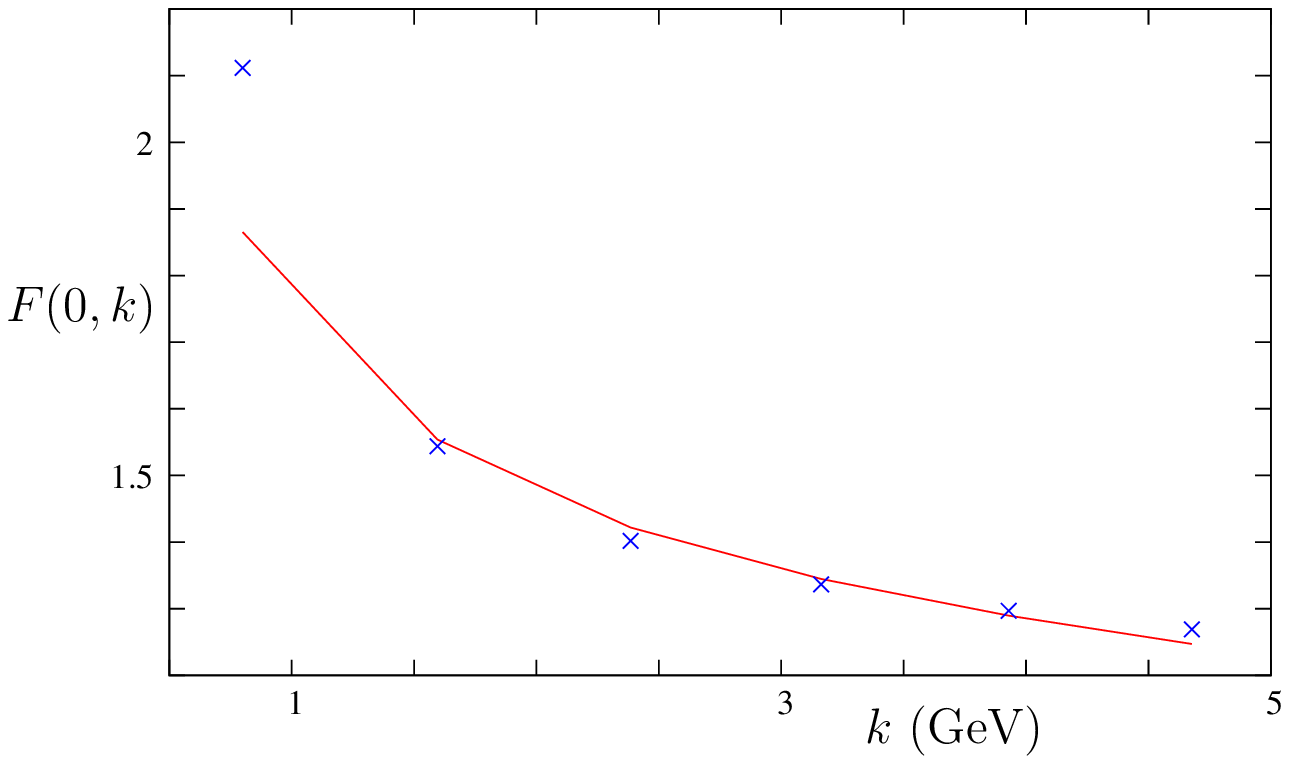}
  \caption{Longitudinal gluon propagator (upper), transverse gluon
    propagator (middle) and ghost dressing function for SU(3) at the
    critical temperature $T_c\simeq 298$ MeV, as a function of momentum. The lattice data
    of \cite{Aouane:2011fv} (crosses) are compared with our one-loop
    calculation (plain line). The renormalization-group scale is chosen to be
    $1$~GeV. The best fit is obtained for $g=3.2$ and $m=300$~MeV.}
  \label{fig_fits_su3}
\end{figure}

We compared our results with two sets of lattice data. We used the
lattice data of \cite{Aouane:2011fv}, obtained for SU(3). The lattice
spacing was taken to be rather small so that the first nonvanishing
momentum corresponds to 800~MeV. When comparing these data with our
one-loop calculation, the fitting parameter $m=300$ MeV, $g=3.2$,
defined at the renormalization scale $\mu=1$ GeV, leads to satisfactory
agreement for the gluon propagator (longitudinal and transverse parts)
as well as for the ghost dressing function, for all accessible
temperatures. This is represented in Fig.~\ref{fig_fits_su3} for
$T=T_c\simeq 298$ MeV. For all other available temperatures, we find
fits of the same quality (not shown). 

\begin{figure}[ht]
  \centering
\includegraphics[width=.9\linewidth]{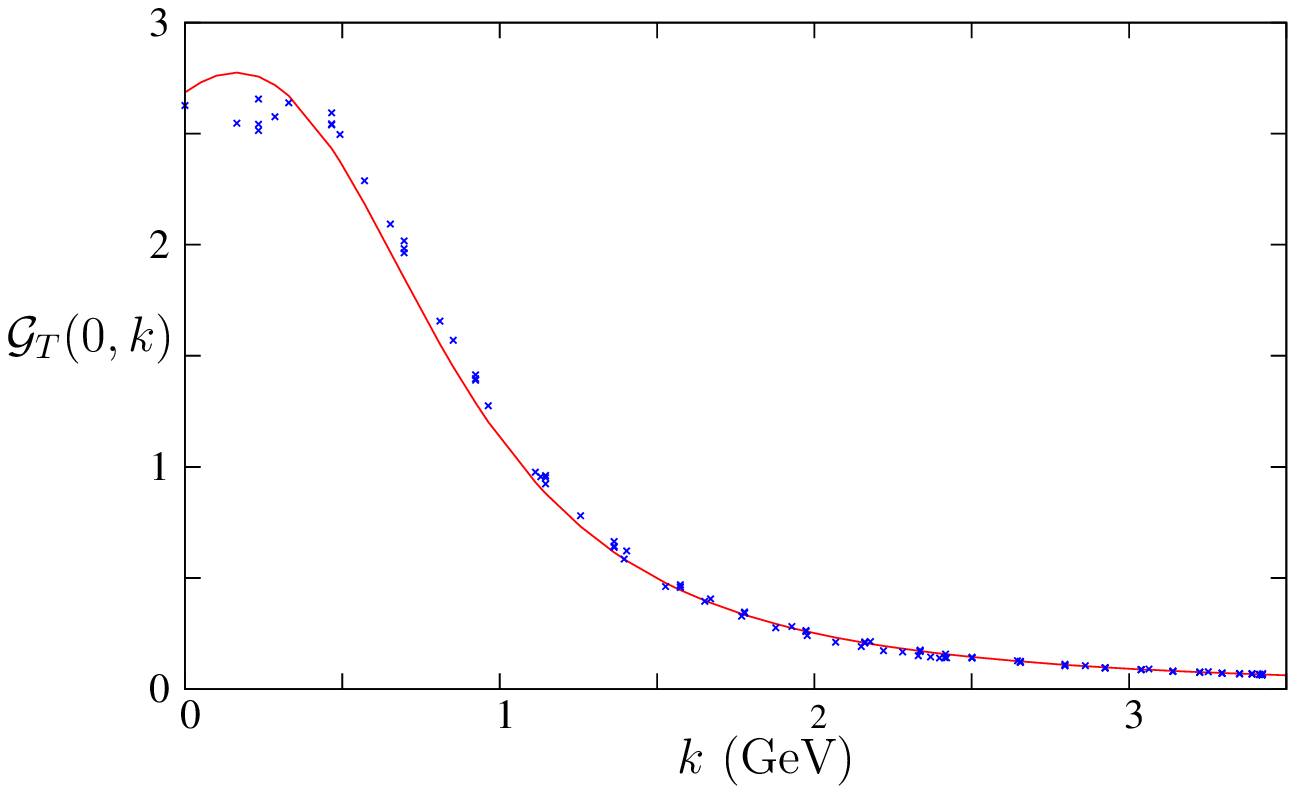}  

\includegraphics[width=.9\linewidth]{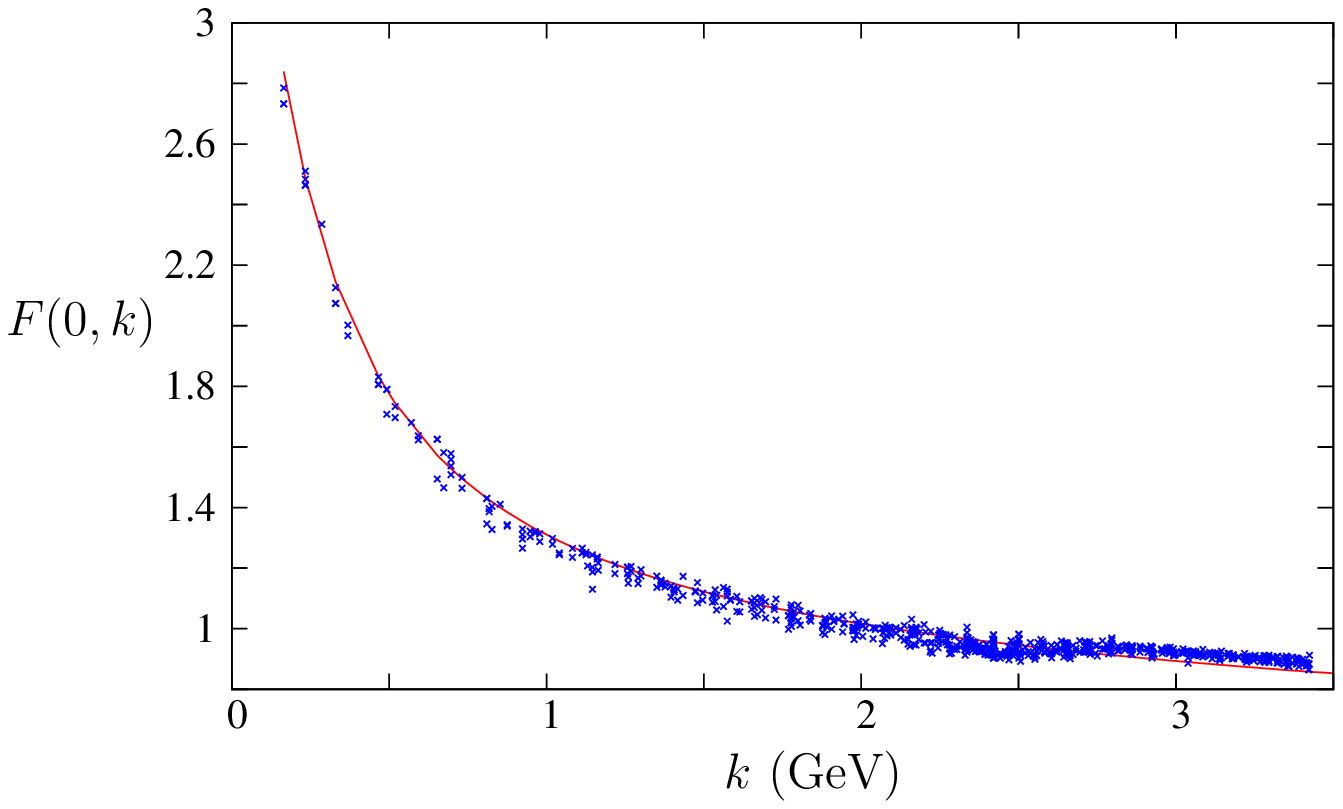}  
  \caption{Transverse gluon propagator (upper) and ghost dressing function (lower) as a function of momentum for $T=T_c$. The lattice data of \cite{Maas:2011ez} are compared with our one-loop calculation. For this temperature, the best fit is obtained for $g=5$, $m=660$ MeV, $Z_A=1.7$ and $Z_c=1.3$.}
  \label{fig_su2}
\end{figure}

\begin{figure}[ht]
  \centering
  \includegraphics[width=.9\linewidth]{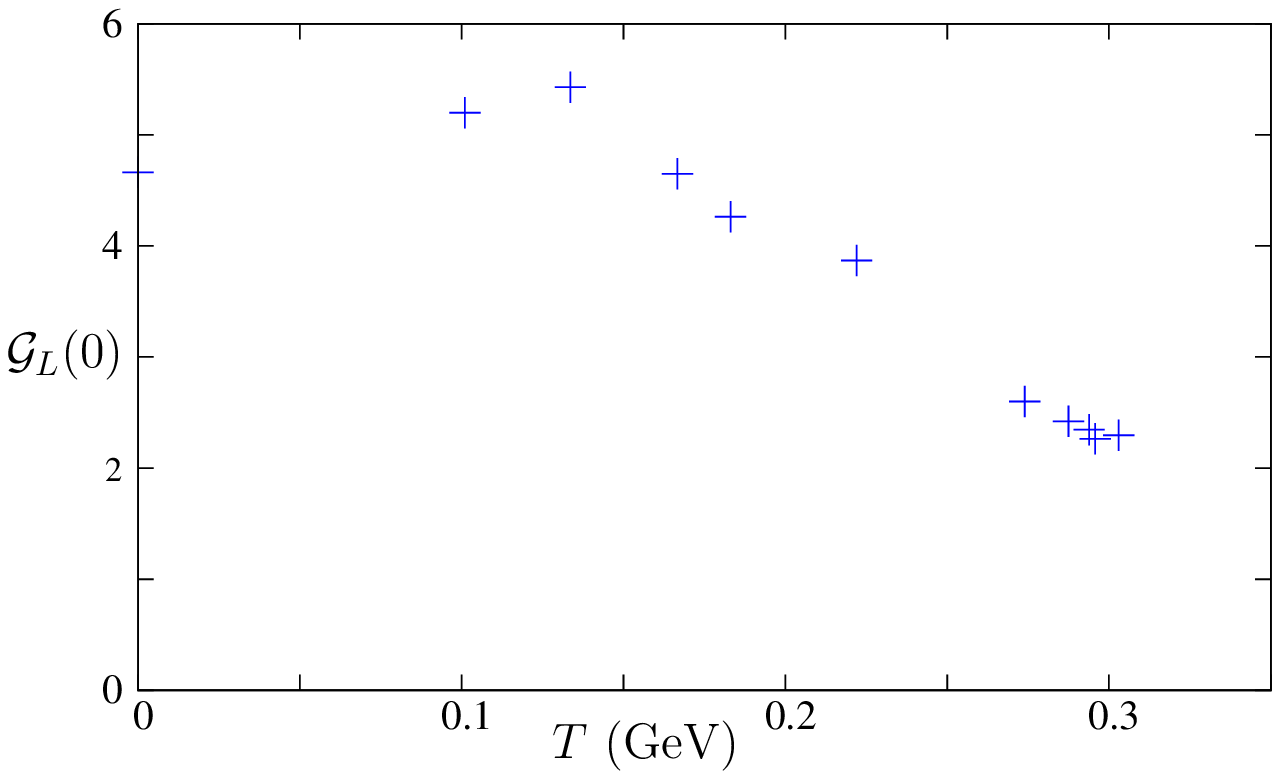}
  \includegraphics[width=.9\linewidth]{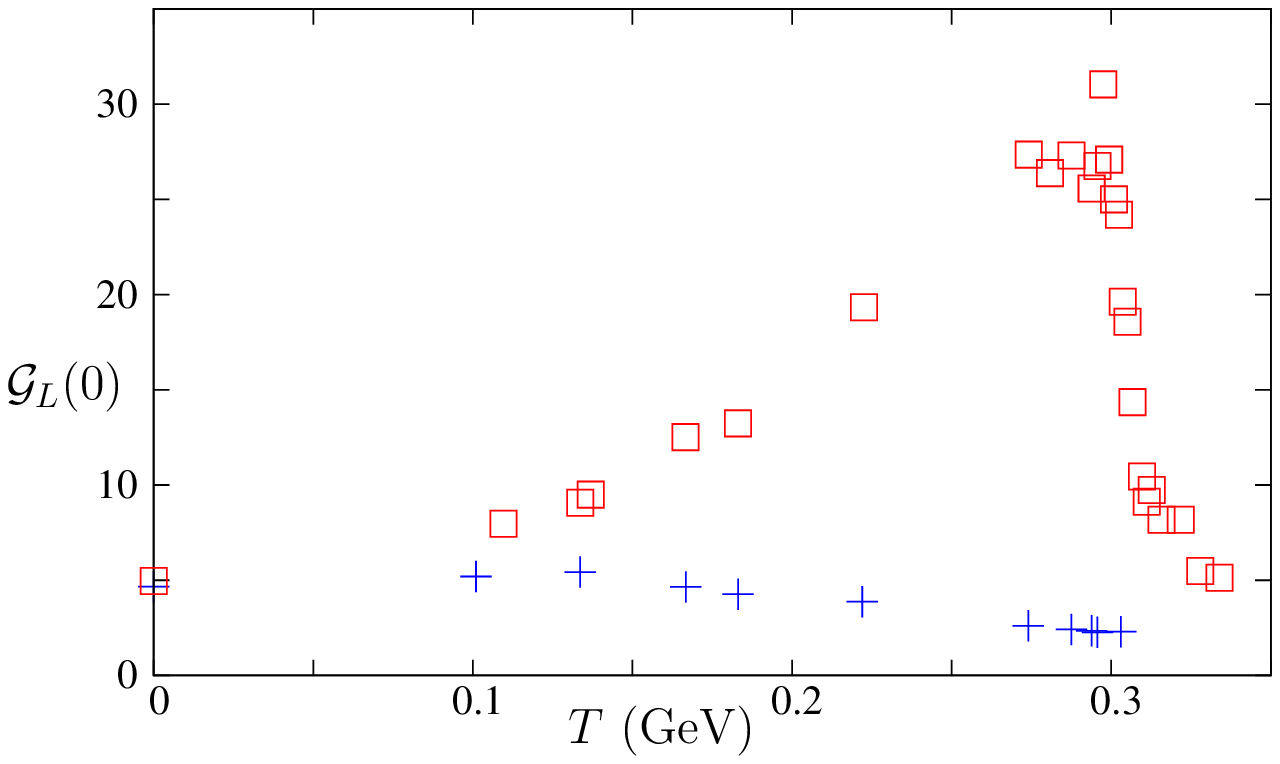}
  \caption{Longitudinal gluon propagator at vanishing momentum as a
    function of temperature obtained by using the fitting parameters
    (crosses, blue). In the lower panel, the squares
    (red) are the lattice data of \cite{Maas:2011ez}.}
  \label{fig_debye_su2}
\end{figure}

We also compared our results with those of \cite{Maas:2011ez} which
are obtained for the SU(2) group and which reach lower energies. As
already discussed, the gluon transverse propagator seems to have
better convergence properties than the longitudinal one (see
\cite{Cucchieri11, Cucchieri12}). For each temperature, we therefore
looked for the set of parameters ($Z_A$, $Z_c$, $g$ and $m$) that lead
\begin{figure}[ht]
  \centering
\includegraphics[width=.83\linewidth]{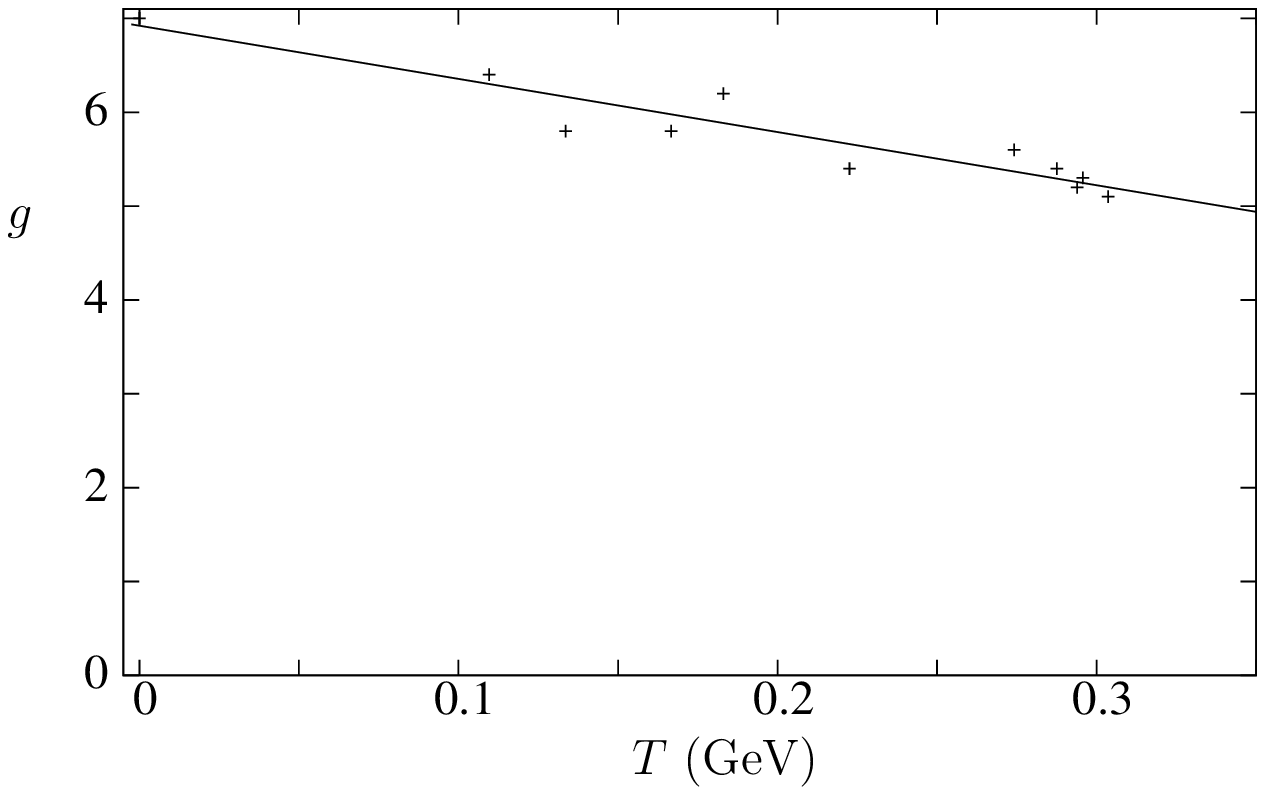}  

\hspace{-.5cm}\includegraphics[width=.9\linewidth]{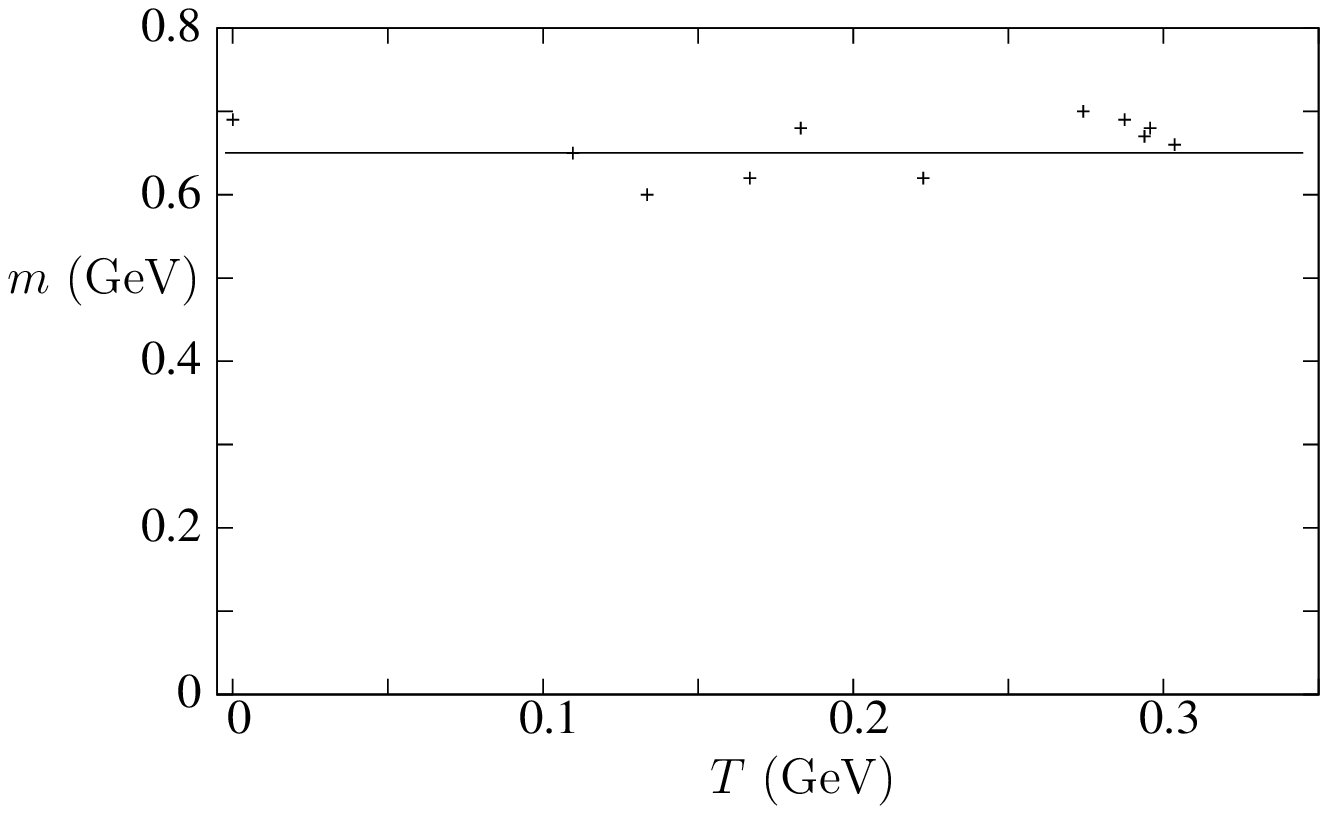}  
  \caption{Variation of the fitting parameters $g$ (top) and $m$ (bottom) with the temperature. The lines are linear fits.}
  \label{fig_gmT}
\end{figure}
to the best agreement between the lattice data and our one-loop
calculation. {  In practice, we find that varying the fitting parameters of roughly 10\% around the optimal value does not modify significantly the quality of the fits; see \cite{Pelaez:2013cpa} for a detailed discussion of this phenomenon at zero temperature.} As shown in Fig.~\ref{fig_gmT}, $m$ is roughly
independent of the temperature, with $m\simeq 650$ MeV, while $g$ decreases from 7 at $T=0$ to 5 at $T=T_c$. {  The fluctuations seen in Fig.~\ref{fig_gmT} can be attributed to the fact mentioned above that we obtain fits of the same quality when changing slightly the values of fitting parameters.} The field
renormalization $Z_A$ decreases from 2.2 to 1.8 in the same
temperature range. The agreement with the transverse gluon propagator
and ghost dressing function is satisfying. This is exemplified in
Fig.~\ref{fig_su2} where $T=T_c$. We found fits of the same quality
for the other available temperatures (not shown). In particular, the
divergence of the ghost dressing function that was observed in
Fig.\ref{fig_ghost_dressing} is not present anymore because of the
decrease of the coupling constant with the temperature.\footnote{The need for a proper running of the coupling constant with the temperature in order to maintain a temperature-independent ghost propagator has also been emphasized in \cite{Fister11,Fukushima:2013xsa,Huber:2013yqa}.}
We then used these parameters to determine the behavior of the
longitudinal propagator at vanishing momentum and the result is
depicted in Fig. \ref{fig_debye_su2}. The nonmonotonous behavior observed in Fig.~\ref{fig_debye} is reproduced. The position of the peak is again determined by \Eqn{eq:peak}, which corresponds roughly to $T_c/2$. The ratio between the zero-temperature and the minimum values of the Debye mass is roughly 1.1. We thus obtain a  qualitative, although not quantitative, agreement for this quantity.

%%%%%
\section{Conclusions}
\label{sec_concl}

We have performed a one-loop calculation of the ghost and gluon two-point correlators at finite temperature in the class of Landau gauges proposed in \cite{Serreau:2012cg}, which is perturbatively equivalent to the Curci-Ferrari model in the Landau limit. Perturbation theory in this framework does very well in describing Yang-Mills correlators at zero temperature, showing that the dominant infrared physics is well described by a simple effective gluon mass. The object of the present work is to investigate to what extent this observation remains true at finite temperature, where one expects specific temperature-dependent infrared effects, in particular, close to the critical temperature. Our results show that both the ghost and the magnetic gluon propagators can be accurately described by a simple mass term also at finite temperature. In contrast, the nonmonotonous behavior of the electric gluon propagator in the infrared is poorly described in the transition region, although we do observe a slight dip of the Debye mass.\footnote{One should keep in mind though that the rapid variation observed in lattice calculations is still poorly under control and shows a strong sensitivity to discretization artifacts \cite{Cucchieri11,Cucchieri12,Mendes:2014gva}.} {  We mention that our results seem to agree qualitatively with those of  \cite{Fister11} although a direct comparison is difficult because these authors do not plot the temperature dependence of the Debye mass.}

The present results are consistent with the view that the electric gluon correlator and, in particular, the Debye mass might be sensitive to the existence of long range correlations near the transition temperature \cite{Aouane:2011fv,Maas:2011ez,Xu:2011ud}. Near a second order or weak first order phase transition, one expects critical fluctuations of some order parameter to play an important role. The latter are usually not captured by a perturbative calculation as performed here and require another ingredient related to the presence of a nontrivial order parameter. An interesting extension of the present study is to account for a possible nontrivial background gluon field, possibly related to the Polyakov loop \cite{Braun:2007bx,Marhauser:2008fz,Fister:2013bh}. This is work in progress.

{  Finally, we mention that the electric-magnetic asymmetry of the $A^2$ condensate has been proposed as a sensitive probe of the deconfinement transition in the SU($2$) theory in \cite{Chernodub:2008kf}. It is easy to compute this asymmetry in the present framework at leading order, which uses the tree-level propagator \eqn{eq:freeprop}. This is unable to reproduce the nonmonotonous behavior as a function of $T$ observed in lattice data  \cite{Chernodub:2008kf}. It would be of great interest to check whether the next-to-leading order contribution, which involves the one-loop gluon propagators computed here, possibly including a nontrivial background gluon field, would modify this. We leave this for future investigation.}

%%%%%
\section*{Acknowledgements}

{  We thank L. Fister for interesting discussions as well as} A.~Maas, M.~M\"uller-Preussker and O.~Oliveira for providing their numerical data. M.~T. and N.~W. acknowledge partial support from the PEDECIBA and ECOS programs. N.~W. thanks the LPTMC (UPMC) and IPhT (CEA, Saclay) on 2012-2013 and support from the European Research Council under the Advanced Investigator Grant No. ERC-AD-267258 (QCDMat).

\appendix

%%%%%
\section{Thermal contributions}
\label{appsec:thermal}

In the present section we detail the calculation of the thermal parts of the loop integrals \eqn{eq:int0}-\eqn{eq:int1} and \eqn{appeq:ITdef}-\eqn{appeq:ILdef}. We note that, because these contributions are finite, we can set $\epsilon\to0$. 

%%%
\subsection{Matsubara sums}
\label{appsec_Matsusum}

We now proceed to the explicit evaluation of the Matsubara sums and momentum integrals in Eqs. \eqn{eq:int0}, \eqn{eq:int1}, \eqn{appeq:ITdef} and \eqn{appeq:ILdef}. To this purpose we note that if $f(z)$
 is a complex function with poles away from the Matsubara frequencies and that decreases fast enough as $|z|\to\infty$, we have
\beq
T\sum_n f(i\omega_n)=-\sum_{\mbox{\scriptsize p\^oles of f}} n(\omega)\,{\rm Res}\,f|_\omega\,,
\eeq
where $n_x=1/(\exp(\beta x)-1)$ is the Bose-Einstein distribution function. In this way we obtain
\beq\label{eq:tad}
J^\alpha=\int_{\vec{q}}\frac{n_{\varepsilon_{\alpha,q}}-n_{-\varepsilon_{\alpha,q}}}{2\varepsilon_{\alpha,q}}=\int_{\vec{q}}\frac{1+2n_{\varepsilon_{\alpha,q}}}{2\varepsilon_{\alpha,q}},
\eeq
where $\varepsilon_{\alpha,q}\equiv\sqrt{q^2+\alpha^2}$ and we have used $n_{-x}=-1-n_x$. The term 1 in the numerator of (\ref{eq:tad}) corresponds to the vacuum contribution and will be determined in the next section by other methods. If we focus only on the thermal contributions, we can write
\beq
J^\alpha_{\rm th}=\int_{\vec{q}}\frac{n_{\varepsilon_{\alpha,q}}}{\varepsilon_{\alpha,q}}.
\eeq
Similar considerations lead to 
\begin{align}
\label{appeq:bub0}
I^{\alpha\beta}_{\rm th}(K)&= \int_{\vec q}\,\frac{n_{\varepsilon_{\alpha,q}}}{\varepsilon_{\alpha,q}}\,{\rm Re}\,G_\beta(\omega+i\varepsilon_{\alpha,q},l)\nn&+ \int_{\vec q}\,\frac{n_{\varepsilon_{\beta,q}}}{\varepsilon_{\beta,q}}\,{\rm Re}\,G_\alpha(\omega+i\varepsilon_{\beta,q},l),
\end{align}
where we have performed the change of variables $\vec{q}\to\vec{k}-\vec{q}$ in the second line. Similarly 
\begin{align}
\label{appeq:bub1}
&2I_{T,{\rm th}}^{\alpha\beta}(K)\nn&=\int_{\vec q}\,\left(q^2-\frac{(\vec{k}\cdot\vec{q})^2}{k^2}\right)\,\frac{n_{\varepsilon_{\alpha,q}}}{\varepsilon_{\alpha,q}}\,{\rm Re}\,G_\beta(\omega+i\varepsilon_{\alpha,q},l)\nonumber\\
& +\int_{\vec q}\,\left(q^2-\frac{(\vec{k}\cdot\vec{q})^2}{k^2}\right)\,\frac{n_{\varepsilon_{\beta,q}}}{\varepsilon_{\beta,q}}\,{\rm Re}\,G_\alpha(\omega+i\varepsilon_{\beta,q},l).\nn
\end{align}
where we used the fact that 
\beq
P^{T}_{\mu\nu}(K)Q_\mu Q_\nu=q^2-\frac{(\vec{k}\cdot\vec{q})^2}{k^2}
\eeq
is invariant under the change of variables $\vec{q}\to\vec{k}-\vec{q}$. Finally
\begin{align}
\label{appeq:bub2}
&k^2K^2I_{L,{\rm th}}^{\alpha\beta}(K)\nn&=\int_{\vec q}\,\frac{n_{\varepsilon_{\alpha,q}}}{\varepsilon_{\alpha,q}}\,{\rm Re}\,\left[\left(ik^2\varepsilon_{\alpha,q}+\omega(\vec{k}\cdot\vec{q})\right)^2\!G_\beta(\omega+i\varepsilon_{\alpha,q},l)\right]\nonumber\\
&+\int_{\vec q}\,\frac{n_{\varepsilon_{\beta,q}}}{\varepsilon_{\beta,q}}\,{\rm Re}\,\left[\left(ik^2\varepsilon_{\beta,q}+\omega(\vec{k}\cdot\vec{q})\right)^2\!G_\alpha(\omega+i\varepsilon_{\beta,q},l)\right],\nn
\end{align}
where we have used
\beq
P^{L}_{\mu\nu}(K)Q_\mu Q_\nu=\frac{1}{k^2K^2}\Big(\omega_nk^2-\omega(\vec{k}\cdot\vec{q})\Big)^2.
\eeq

%%%
\subsection{Angular integrations}
\label{appsec_angular}

All angular integrations can be explicitly performed in the expressions derived in the previous section, leaving one-dimensional radial integrals to be performed numerically. The thermal part of the tadpole integral reads
\beq
J_{\rm th}^\alpha=\frac{1}{2\pi^2}\int_0^\infty dq\,q^2\,\frac{n_{\varepsilon_{\alpha,q}}}{\varepsilon_{\alpha,q}}\,.
\eeq
The angular integrations in Eqs. \eqn{appeq:bub0}, \eqn{appeq:bub1} and \eqn{appeq:bub2} involve the integrals
\beq
 I_n=\int_{-1}^1\frac{d\cos\theta\,\cos^n\theta}{A+\cos\theta}
\eeq
for $n=0,1,2$. The latter are given by
\beq
 I_0= \ln\frac{A+1}{A-1}\,,\quad I_1 =2-AI_0\quad{\rm and}\quad I_2 =  -AI_1.
\eeq
After some simple algebra the thermal contributions to the relevant loop integrals can be cast in the following forms:
\begin{widetext}
\bea
 I^{\alpha\beta}_{\rm th}(K) &=&-\frac{1}{8\pi^2k}\int_0^\infty \!\!dq\,q\,\frac{n_{\varepsilon_{\alpha,q}}}{\varepsilon_{\alpha,q}}\,{\rm Re}\, g_\beta(\omega+i\varepsilon_{\alpha,q};q,K)+(\alpha\leftrightarrow\beta),\\
  I_{T,{\rm th}}^{\alpha\beta}(K)&=&\frac{1}{64\pi^2k^3}\!\int_0^\infty \!\!dq\,q\,\frac{n_{\varepsilon_{\alpha,q}}}{\varepsilon_{\alpha,q}}\!\left[4kq\left(K^2+\beta^2-\alpha^2\right)+{\rm Re}\,\ell_T^\beta(\omega+i\varepsilon_{\alpha,q};q,K)\right]+(\alpha\leftrightarrow\beta), \nn\\
  I_{L,{\rm th}}^{\alpha\beta}(K)&=&-\frac{\omega^2}{32\pi^2K^2k^3}\!\int_0^\infty\!\!dq\,q\,\frac{n_{\varepsilon_{\alpha,q}}}{\varepsilon_{\alpha,q}}\!\left[4kq\left(K^2+\beta^2-\alpha^2\right)+{\rm Re}\,\ell_L^\beta(\omega+i\varepsilon_{\alpha,q};q,K)\right]\nn&+&(\alpha\leftrightarrow\beta),
\eea
\end{widetext}
where we introduced the functions
\beq
 g_\beta(z;q,K)=\ln\frac{z^2+\varepsilon^2_{\beta,k-q}}{z^2+\varepsilon^2_{\beta,k+q}},
\eeq
\beq
 \ell_T^\beta(z;q,K)=\left(\varepsilon_{\beta,k+q}^2+z^2\right)\left(\varepsilon_{\beta,k-q}^2+z^2\right)\,g_\beta(z;q,K)
\eeq
and
\beq
 \ell_L^\beta(z;q,K)=\left[\varepsilon_{\beta,q}^2+z^2+k^2\left({2z\over\omega}-1\right)\right]^{2}\!g_\beta(z;q,K).
\eeq
We shall also make use of the following expansion:
\bea
 I^{\alpha\beta}_{\rm th}(0,k) & = & \frac{1}{2\pi^2}\int_0^\infty dq\,q^2\,\frac{n_{\varepsilon_{\alpha,q}}}{\varepsilon_{\alpha,q}}\,\,\frac{1}{\beta^2-\alpha^2}\nn
 & - & \frac{k^2}{2\pi^2}\int_0^\infty dq\,q^2\,\,\frac{n_{\varepsilon_{\alpha,q}}}{\varepsilon_{\alpha,q}}\,\frac{1}{(\beta^2-\alpha^2)^2}\nn
 & + & \frac{2k^2}{3\pi^2}\int_0^\infty dq\,q^4\,\,\frac{n_{\varepsilon_{\alpha,q}}}{\varepsilon_{\alpha,q}}\,\frac{1}{(\beta^2-\alpha^2)^3}\nn
 & + & (\alpha\leftrightarrow\beta)+{\cal O}(k^4)
\eea
as well as the values
\begin{align}
&I_{T,{\rm th}}^{\alpha\beta}(0,k\to 0)\nn&=\frac{1}{6\pi^2}\int_0^\infty dq\,q^4\,\left(\frac{n_{\varepsilon_{\alpha,q}}}{\varepsilon_{\alpha,q}}-\frac{n_{\varepsilon_{\beta,q}}}{\varepsilon_{\beta,q}}\right)\,\,\frac{1}{\beta^2-\alpha^2},\\
&I_{L,{\rm th}}^{\alpha\beta}(0,k\to 0)\nn& =-\frac{1}{2\pi^2}\int_0^\infty\!\!dq\,q^2\,\Big(n_{\varepsilon_{\alpha,q}}\varepsilon_{\alpha,q}-n_{\varepsilon_{\beta,q}}\varepsilon_{\beta,q}\Big)\,\,\frac{1}{\beta^2-\alpha^2}.
\end{align}
These last three expressions can be obtained more directly by noting, from (\ref{eq:int1}), (\ref{appeq:ITdef}) and (\ref{appeq:ILdef}), that
\bea
I^{\alpha\beta}(0,k\to 0) & = & \int_Q G_\alpha(Q)G_\beta(Q),\\
\left.\frac{\partial^2 I^{\alpha\beta}}{\partial k_i\partial k_j}\right|_{0,k\to 0} & = & \left[-\int_Q G_\alpha(Q)G^2_\beta(Q)\right.\nn
& & \hspace{0.1cm}\left.+\frac{4}{3}\int_Q q^2G_\alpha(Q)G^3_\beta(Q)\right]\delta_{ij},\\
I^{\alpha\beta}_T(0,k\to 0) & = & \frac{1}{d-1}\int_Q q^2 G_\alpha(Q)G_\beta(Q),\\
I^{\alpha\beta}_L(0,k\to 0) & = & J^\beta-\int_Q (q^2+\alpha^2) G_\alpha(Q)G_\beta(Q),\nn
\eea
and by using (\ref{eq:id1}) and the identity
\begin{align}
\int_Q q^{2p}G^n_\alpha(Q) &= -\frac{1}{n-1}\int_Q q^{2p}\frac{dG^{n-1}_\alpha(Q)}{dq^2} \nn&\,\,\hat{=} \,\,\frac{2p+1}{2(n-1)}\int_Q q^{2p-2}G^{n-1}_\alpha(Q),\label{eq:ibp}
\end{align}
to bring all the integrals in the form $\int_Q q^{2p}G_\alpha(Q)$. Strictly speaking the second equality in (\ref{eq:ibp}) applies to the thermal parts of the integrals as emphasized by the the symbol $\hat{=}$.

%%%%%
\section{Vacuum contributions}
\label{appsec:vac}

Vacuum contributions can be obtained from the above section by keeping the discarded nonthermal parts. Equivalently, they can be computed in a standard way by replacing Matsubara sums by frequency integrals in Eqs. \eqn{eq:int0}-\eqn{eq:int1} and \eqn{appeq:ITdef}-\eqn{appeq:ILdef}, that is by replacing $\int_Q\to\int\frac{d^dq}{(2\pi)^d}$. We follow the latter approach here. 

The vacuum one-loop integrals are UV divergent and must be regularized. We employ dimensional regularization and define $\bar\mu^2\equiv 4\pi\mu^2 e^{-\gamma}$.
The tadpole loop reads
\beq
J_{\rm vac}^\alpha =\frac{\mu^{2\epsilon}\Gamma\left(1-{d\over2}\right)}{(4\pi)^{d\over2}}(\alpha^2)^{{d\over2}-1} \,\,\tilde=\,\,-\frac{\alpha^2}{16\pi^2}\left(\frac{1}{\epsilon}+\ln\frac{\bar\mu^2}{\alpha^2}+1\right).
\eeq
where the symbol $\tilde =$ means that we discard terms of relative order ${\cal O}(\epsilon)$. The integral \eqn{eq:int1} is obtained by introducing a Feynman parameter, leading to
\bea
 I_{\rm vac}^{\alpha\beta}(K) &=&\frac{\mu^{2\epsilon}\Gamma\left(\epsilon\right)}{(4\pi)^{d\over2}}\int_0^1 \!\frac{dx }{\left[x\alpha^2+(1-x)\beta^2+x(1-x) K^2\right]^{\epsilon}}\nn
&\tilde=&\frac{1}{16\pi^2}\left\{\frac{1}{\epsilon}+2+\ln\frac{\bar\mu^2}{K^2}-\frac{1}{2}\ln\!\left(C^2_{\alpha\beta}(K)-{1\over4}\right)\right.\nn
&+&\left.C_{\alpha\beta}(K)\ln\frac{C_{\alpha\beta}(K)-{1\over2}}{C_{\alpha\beta}(K)+{1\over2}}\right\}+(\alpha\leftrightarrow\beta),
\eea
where we defined
\beq
 C_{\alpha\beta}(K)=\frac{B_{\alpha\beta}(K)+\alpha^2-\beta^2}{2K^2}
\eeq
with (notice that $B_{\alpha\beta}(K)=B_{\beta\alpha}(K)$)
\beq
 B_{\alpha\beta}(K)=\sqrt{K^4+2K^2\left(\alpha^2+\beta^2\right)+\left(\alpha^2-\beta^2\right)^2}.
\eeq

Due to Lorentz symmetry, the vacuum contributions to the transverse and longitudinal projections \eqn{appeq:ITdef}-\eqn{appeq:ILdef} are equal: $I_{T,{\rm vac}}^{\alpha\beta}(K)=I_{L,{\rm vac}}^{\alpha\beta}(K)=I_{\perp,{\rm vac}}^{\alpha\beta}(K)$, with
\beq
\label{appeq:Iperpvac}
 I_{\perp}^{\alpha\beta}(K)=\frac{P_{\mu\nu}^\perp(K)I_{\mu\nu}^{\alpha\beta}(K)}{d-1}.
\eeq
Writing 
\beq
 P_{\mu\nu}^\perp(K)Q_\mu Q_\nu=\frac{Q^2K^2-(Q\cdot K)^2}{K^2}
\eeq
and using similar manipulations as described in Sec. \ref{sec:ghostself}, we can rewrite the projection \eqn{appeq:Iperpvac} in terms of the integrals \eqn{eq:int0} and \eqn{eq:int2}
\begin{align}
(d-1)I_{\perp}^{\alpha\beta}(K) &= \frac{K^2+\alpha^2-\beta^2}{4K^2}J^\beta- {B_{\alpha\beta}^2(K)\over4}I^{\alpha\beta}(K)\nn&+(\alpha\leftrightarrow\beta)\,.
\end{align}

%%%%%
\section{Gluon self-energy}
\label{appsec:sketch}

Here, we show how the expression of the gluon self-energy, represented on Fig. \ref{fig:gluon}, can be reduced to the elementary integrals \eqn{eq:int0}, \eqn{eq:int1} and \eqn{eq:int2}.

The contribution from the tadpole diagram in Fig. \ref{fig:gluon} reads 
\begin{align}
\Pi^{\rm tad}_{\mu\nu}&=\int_Q\big[\delta_{\mu\nu} P^\perp_{\rho\rho}(Q)-P^\perp_{\mu\nu}(Q)\big]G_m(Q)\nn&=(d-2)\delta_{\mu\nu} J^m+I^{m0}_{\mu\nu}(0),
\end{align}

where we wrote $P^\perp_{\mu\nu}(Q)=Q_\mu Q_\nu G_0(Q)$ in writing the second line.

The ghost loop contribution reads
\begin{align}
\Pi^{\rm gh}_{\mu\nu}(K)&=-{1\over2}\int_Q \big[Q_\mu L_\nu+L_\mu Q_\nu\big] G_0(Q)G_0(L)\nn&=I^{00}_{\mu\nu}(K)-\frac{1}{2}K_\mu K_\nu I^{00}(K),
\end{align}
where we used 
\beq 
\label{appeq:simplif}
Q_\mu L_\nu+L_\mu Q_\nu=K_\mu K_\nu-Q_\mu Q_\nu-L_\mu L_\nu
\eeq
in writing the second line.

The gluon loop contribution is the most involved. It reads
\begin{widetext}
\bea
\Pi^{\rm gl}_{\mu\nu}(K) & = & -\frac{1}{2}\int_Q (L-Q)_\mu(L-Q)_\nu \,{\rm tr}\,\Big[P^\perp(Q)P^\perp(L)\Big]G_m(Q)G_m(L)\nonumber\\
& - & \int_Q \Big[(K+Q)\cdot P^\perp(L)\cdot (K+Q)\Big] P^\perp_{\mu\nu}(Q)\,G_m(Q)G_m(L)\nonumber\\
& + & \int_Q \Big[(K+L)\cdot P^\perp(Q)\Big]_\mu \Big[(K+Q)\cdot P^\perp(L)\Big]_\nu G_m(Q)G_m(L)\nn
\label{appeq:Pigl}
& - & \int_Q (L-Q)_\mu \Big[(K+Q)\cdot P^\perp(L)\cdot P^\perp(Q)\Big]_\nu G_m(Q)G_m(L)+(\mu\leftrightarrow \nu),
\eea
\end{widetext}
where the symmetrization $(\mu\leftrightarrow\nu)$ applies only to the last line, the other ones being explicitly symmetric.

We show in detail how to treat the contribution on the first line, which we denote as $\Pi_{\mu\nu}^{\rm gl,(1)}(K)$ in the following. The trace reads
\beq
{\rm tr}\,\big[P_\perp(Q)\cdot P_\perp(L)\big]=(d-2)+\frac{(Q\cdot L)^2}{Q^2L^2}.
\eeq
The contribution from the term $\propto d-2$ can be directly reduced to simple integrals by using again \Eqn{appeq:simplif}:
\beq
 (L-Q)_\mu(L-Q)_\nu=2(Q_\mu Q_\nu+L_\mu L_\nu)-K_\mu K_\nu.
\eeq
One gets
\bea
&&\hspace{-.7cm}\Pi_{\mu\nu}^{\rm gl,(1)}(K) = -\frac{d-2}{2}\Big[4\, I^{mm}_{\mu\nu}(K)-K_\mu K_\nu\,I^{mm}(K)\Big]\nonumber\\
\label{appeq:Pi1}
& &\hspace{-.7cm}-  \frac{1}{2}\int_Q (L-Q)_\mu(L-Q)_\nu \frac{(Q\cdot L)^2}{Q^2L^2}\,G_m(Q)G_m(L).
\eea
The remaining integral can be simplified using similar manipulations as in the discussion of the ghost self-energy in Sec. \ref{sec:ghostself}. We first eliminate the $1/Q^2L^2$ term by using \Eqn{eq:id1}. We are left with products of propagators $G_\alpha(Q) G_\beta(L)$ with $\alpha,\beta=m,0$. Now, writing
\beq
 2Q\cdot L = K^2+\alpha^2+\beta^2-(Q^2+\alpha^2)-(L^2+\beta^2),
\eeq
we have the identity
\begin{align}
2 (Q\cdot L)\,G_\alpha(Q)G_\beta(L)&=\left(K^2+\alpha^2+\beta^2\right)G_\alpha(Q)G_\beta(L)\nn&-G_\alpha(Q)-G_\beta(L).
\end{align}
Applying this trick twice, we get
\begin{widetext}
\bea
\frac{(Q\cdot L)^2}{Q^2L^2}G_m(Q)G_m(L)& =&  \frac{(K^2+2m^2)^2}{4m^4} G_m(Q)G_m(L)+\frac{K^4}{4m^4} G_0(Q)G_0(L)\nonumber\\
& -& \frac{(K^2+m^2)^2}{4m^4}\Big[G_m(Q)G_0(L)+G_0(Q)G_m(L)\Big]\nonumber\\
\label{appeq:decomp}
& +&\frac{1}{4}\Big[G_0(Q)G_m(Q)+G_0(L)G_m(L)\Big],
\eea
where we used again \Eqn{eq:id1} in the last line. The above expression is such that when plugged in \Eqn{appeq:Pi1}, each term gives a well-defined sum-integral in dimensional regularization.
Inserting \eqn{appeq:decomp} in \eqn{appeq:Pi1}, we finally get
\bea
 \hspace{-.8cm}\Pi^{\rm gl,(1)}_{\mu\nu}(K) &=&  \frac{(K^2+m^2)^2}{4m^4} \Big[2\left(I^{m0}_{\mu\nu}(K)+ I^{0m}_{\mu\nu}(K)\right)-K_\mu K_\nu\,I^{m0}(K)\Big]\nn
&-&\left({d-2\over2}+\frac{(K^2+2m^2)^2}{8m^4}\right)\Big[4\, I^{mm}_{\mu\nu}(K)-K_\mu K_\nu\,I^{mm}(K)\Big]\nn
&-&\frac{(K^2)^2}{8m^4} \Big[4\, I^{00}_{\mu\nu}(K)-K_\mu K_\nu\,I^{00}(K)\Big]- \frac{1}{4}\Big[4\, I^{m0}_{\mu\nu}(0)+K_\mu K_\nu\, I^{m0}(0)\Big].\,
\eea

The remaining three contributions corresponding to the three last lines in \Eqn{appeq:Pigl} can be treated along similar lines. After a straightforward calculation, we obtain, for the gluon polarization tensor at one loop:
\bea
&&\Pi_{\mu\nu}(K)= \nn
&&\left(1-\frac{K^4}{2m^4}\right)I^{00}_{\mu\nu}(K)+  \left(1+\frac{K^2}{m^2}\right)^{\!\!2}\frac{I_{{\mu\nu}}^{m0}(K)+I_{{\mu\nu}}^{0m}(K)}{2}-2\left[d-2+\left(1+\frac{K^2}{2m^2}\right)^{\!\!2}\right]I^{mm}_{\mu\nu}(K)\nn
& &+ \delta_{\mu\nu}\left[(d-2) J^m-(K^2+m^2)\,I^{m0}(0)+\frac{(K^2+m^2)^2}{m^2}\, I^{m0}(K)-K^2\left(4+\frac{K^2}{m^2}\right) I^{mm}(K)\right]\nn
& &+ K_\mu K_\nu\left[\left(\frac{K^4}{8m^4}-\frac{1}{2}\right) I^{00}(K)-\,\frac{(K^2+m^2)(K^2+5m^2)}{4m^4}  I^{m0}(K)-\frac{1}{4}I^{m0}(0)\right]\nn
& &+K_\mu K_\nu\left[\frac{d-2}{2}+\frac{(K^2+6m^2)^2}{8m^4}\right] I^{mm}(K)+ \frac{K^2+m^2}{m^2}\Big[I_{{\mu\nu}}^{m0}(K)-I_{{\mu\nu}}^{0m}(K)\Big].
\eea
Here, we have organized the terms in such a way that the last two lines do not contribute to the projections $\Pi^{T,L}(K)$ thanks to the properties $K_\mu P^{T,L}_{\mu\nu}(K)=0$ and \eqn{eq:symmetry}. The result quoted in the main text, \Eqn{eq:gluonself}, follows.
\end{widetext}

%%%%%

\end{document}